\documentclass[a4paper,11pt]{article}
\pdfoutput=1

\usepackage{jheppub}
\usepackage[T1]{fontenc}
\usepackage{tensor}
\usepackage{subfig}
\hypersetup{bookmarksnumbered}
\title{\boldmath The Abrikosov Vortex in Curved Space}

\author[a,b]{Jan Albert}

\affiliation[a]{Simons Center for Geometry and Physics, Stony Brook University,\newline Stony Brook, NY 11794-3636, U.S.A.}

\affiliation[b]{C. N. Yang Institute for Theoretical Physics, Stony Brook University,\newline Stony Brook, NY 11794-3840, U.S.A.}

\emailAdd{jan.albertiglesias@stonybrook.edu}

\abstract{We study the self-gravitating Abrikosov vortex in curved space with and without a (negative) cosmological constant, considering both singular and non-singular solutions with an eye to hairy black holes. In the asymptotically flat case, we find that non-singular vortices round off the singularity of the point particle's metric in 3 dimensions, whereas singular solutions consist of vortices holding a conical singularity at their core. There are no black hole vortex solutions. In the asymptotically AdS case, in addition to these solutions there exist singular solutions containing a BTZ black hole, but they are always hairless. So we find that in contrast with 4-dimensional 't\ Hooft-Polyakov monopoles, which can be regarded as their higher-dimensional analogues, Abrikosov vortices cannot hold a black hole at their core. We also describe the implications of these results in the context of AdS/CFT and propose an interpretation for their CFT dual along the lines of the holographic superconductor.}

\begin{document} 
\maketitle
\flushbottom

\section{Introduction and Summary}
Abrikosov vortices (also known as Nielsen-Olesen vortices) are soliton solutions that occur in the symmetry-broken phase of the $D=2+1$ Abelian-Higgs model. These solutions are the result of a topological map between the $S_{\infty}^1$ at spatial infinity and the $S_{\phi}^1$ of the complex phase of the scalar field and are hence labelled by a topological invariant; the vorticity $N\in \mathbb{Z}$. In addition, they carry a quantized magnetic flux but no electric charge. They were first studied in the context of superconductivity to explain type II superconductors \cite{abrikosov} and later in relation to String Theory \cite{nielsen}. In $D=3+1$, these solutions extend into line solitons that have applications in different contexts: in condensed matter, they describe flux tubes in superconductors, while in cosmology, cosmic strings. But in this work we will stick to the purely 3-dimensional solutions.

One can think of these vortices as the 3-dimensional analogue of $D=3+1$ 't\ Hooft-Polyakov ('tHP) monopoles, which are also topological solutions carrying a quantized magnetic flux \cite{hooft, polyakov}. These monopoles appear in various gauge theories, but perhaps the simplest one is an $SU(2)$ gauge theory with a scalar field in the adjoint representation. They are similar to vortices in that they also live in the symmetry-broken phase of the theory and that they are supported by topology since they are the result of an analogous $S_{\infty}^2\to S_{\phi}^2$ map between spatial infinity and field space. However, an important difference is that monopoles have asymptotically a radial magnetic field of an unbroken $U(1)$ symmetry while no continuous symmetry remains at the edge of Abrikosov vortices.

It has now been known for quite long that these monopoles can hold a small Schwarzschild black hole (BH) within their core when some bounds in parameter space are met \cite{lee,ortiz,forgacs}. Thus, they provide an example of non-trivial matter structure outside the black hole horizon --dubbed ``hair''. This had long been thought impossible by the so-called no-hair theorems, but these only apply to some specific theories and now we know of several examples that evade them \cite{bekenstein}. These objects usually go under the name of {\it hairy magnetic BHs} and they have received increased attention in the recent literature (see e.g.\ \cite{maldacena,bai_2020,mcinnes,bai} and references therein). Given the obvious analogy between vortex and monopole solutions, it seems reasonable to ask whether self-gravitating Abrikosov vortices can also hold a small black hole within their core.

The purpose of this paper is to follow the classical analyses of self-gravitating monopoles \cite{lee,ortiz,forgacs} to answer this question. In general terms, these works first look for non-singular 'tHP monopole solutions in curved space to see the effects that gravity has on the well-known flat space solutions. Then, they insert a small mass at the core of the monopole that seeds a small Schwarzschild BH and they study how the matter fields behave outside its horizon. When the horizon is small enough, they find that two configurations are possible. First, a trivial solution where the matter fields take their asymptotic values already at the horizon, which corresponds to the usual magnetic Reissner–Nordström (RN) black hole; a hairless black hole. Second, a non-trivial solution where the tail of the monopole is still observed outside the horizon, which realizes the hairy magnetic black hole. When comparing their energies, it turns out that the hairy solution is preferred over the hairless one.

Following their steps, we start by studying non-singular self-gravitating Abrikosov vortices in asymptotically (locally) flat space by coupling the Abelian-Higgs model minimally to gravity and making the metric dynamical. These solutions are asymptotic to the metric of a point particle of the same mass as the vortex in $2+1$ dimensions, which is the metric of a cone \cite{deser,clement}. Thus, we find that non-singular vortices round off the conical singularity of the point particle's metric, as we would expect from any massive extended object \cite{peleg}. Turning to singular solutions, we consider the insertion of a point mass at the core of the vortex. In contrast with the higher-dimensional case, in 3-dimensional gravity this just reintroduces a conical singularity there. It does not yield a black hole at all! So it does not make sense to ask if Abrikosov vortices can dress a BH solution when there is no BH to dress.

To go around this issue, we can introduce a negative cosmological constant $\Lambda=-1/\ell^2$ to get asymptotically Anti de Sitter (AdS) space-times, where the Bañados-Teitelboim-Zanelli (BTZ) black hole lives \cite{btz}. In this case, non-singular vortex solutions and vortices holding a conical singularity at their core still exist in certain regions of the parameter space. But now we additionally can consider inserting a BTZ black hole inside the vortex. So the question gets refined to whether Abrikosov vortices can provide hair for a BTZ black hole. The answer is no. We find that they obey a no-hair theorem and so the only possible solution in the presence of a horizon (of any radius) is the one in which the matter fields take their asymptotic values already at the horizon. This corresponds to a hairless BTZ black hole carrying a quantized magnetic flux and it is the lower-dimensional analogue of the magnetic RN black hole described above.

As discussed in \cite{lee}, the reason why 'tHP monopoles evade the no-hair theorems can be tracked down to the effective potential of the theory having a position-dependent absolute minimum. Far away from the origin, the true vacuum is at the symmetry-broken phase, as we would expect, but closer to the origin, the gauge field conspires to return this privilege to the symmetric phase. Thus, the whole monopole is essentially at the true minimum and, even when in the presence of a (small-enough) horizon, it is preferable for the fields around the core to take a monopole shape. However, this feature of the effective potential is intrinsically due to the gauge field being non-Abelian, so it is not present in the Abelian-Higgs model. Only the outskirts of Abrikosov vortices are at the true minimum of the potential; their cores are held solely by the requirement of regularity. As soon as a horizon is present, this condition must be dropped in favor of a condition at the horizon and, as a result, the matter fields relax to their asymptotic values leaving a hairless BTZ black hole.

Being asymptotically AdS, these vortices must have an interpretation in a dual conformal field theory (CFT) by the AdS/CFT correspondence \cite{MAGOO}. The fast asymptotic decay of the matter fields makes vortices normalizable solutions. Therefore, they must be dual to states rather than perturbations of the CFT and, by the state-operator map, they can be associated to an operator of the boundary theory. Given that our solutions are labelled by a topological invariant, we argue that the operator they are dual to is a ``winding operator''. Much like the twist operator \cite{ginsparg}, which is in charge of flipping a sign, this operator is attached to a topological line that makes the fundamental fields wind around the cylinder. Different insertions of this operator place us in specific winding sectors of the Hilbert space.

We also propose an interpretation of the CFT dual on the lines of the holographic superconductor (HS) \cite{horowitz}. In the usual story, a hair/no-hair transition of an electrically charged BH in the bulk is related to a superconducting phase transition at the boundary. In our solutions, in contrast, the BH in the bulk carries a magnetic instead of electric charge. This corresponds to the superconducting phase of the dual theory being modulated by an external magnetic field rather than a chemical potential. In addition, our solutions are labelled by the vorticity, so we expect to see some topological invariant in the boundary theory. These two observations remind us of the famous Little-Parks experiment \cite{LittleParks}, in which they applied a parallel magnetic field to a superconductor in the shape of a cylinder. In this system, the magnetic field induces a supercurrent that winds around the cylinder and can only take quantized values \cite{arutunian}. Thus, we propose that, just like in the HS, our solutions are dual to a superconductor but in the shape of a (1+1)-dimensional cylinder with a coaxial magnetic flux, where the vorticity in the bulk corresponds with the quantized values that the supercurrent can take at the boundary. Despite some drawbacks that we acknowledge in the main text, we find this interpretation promising and we believe it could help us expand our understanding of the holographic superconductor.

This paper has four further sections. In Section \ref{sec:theory}, we review Abrikosov vortices in flat space, couple the theory to gravity and derive the general field equations. But we leave the discussion about boundary conditions for the subsequent sections, where we take particular limits of the general theory. In Section~\ref{sec:flat_section}, we describe the numerical methods used to solve the equations and apply them to asymptotically flat space-times; first in flat space and then with dynamical gravity. As discussed, in $D=2+1$ there is no Schwarzschild BH, so in Section~\ref{sec:ads_section} we consider asymptotically AdS space-times to make contact with the BTZ black hole. We first consider vortices in a fixed AdS$_3$ background and we then make the metric dynamical. We conclude with Section~\ref{sec:ads/cft}, where we discuss our interpretation of the CFT dual of vortex solutions in the abstract and in relation to the holographic superconductor.

\section{Theory}\label{sec:theory}
Consider the Abelian-Higgs model in $D=2+1$ flat space-time, that is, a complex scalar field charged under a $U(1)$ gauge symmetry with a symmetry-breaking potential.
\begin{equation}
	S=\int d^3x \left[-\frac{1}{4}F_{\mu\nu}F^{\mu\nu} + D_\mu \phi\left(D^\mu \phi\right)^* - \frac{\lambda}{4}\left(|\phi|^2-v^2\right)^2\right]\;,
	\label{eq:flataction}
\end{equation}
where the covariant derivative is $D_\mu \phi=\partial_\mu \phi + i e A_\mu \phi$ and $v$ is the vacuum expectation value (vev) that the scalar field acquires under spontaneous symmetry breaking (SSB). This theory admits static soliton solutions that carry magnetic flux but no electric charge known as Abrikosov (or Nielsen-Olesen) vortices \cite{abrikosov,nielsen}.

The usual way of observing these solutions (see e.g.\ \cite{dunne}) is by requiring finiteness of the static (magnetic) energy functional,
\begin{equation}
	\mathcal{E}=\int d^2x\left[\frac{1}{2}B^2+\big|\vec{D}\phi\big|^2+\frac{\lambda}{4}\left(|\phi|^2-v^2\right)^2\right]\;,
	\label{eq:energy_functional}
\end{equation}
where the magnetic field is the pseudoscalar $B=F_{21}$. This immediately implies that the system must be asymptotically in the symmetry-broken phase, with the scalar field $\left|\phi\right|\to v$, but it leaves the freedom of a complex phase. In vortex solutions, this phase winds $N$ times as one moves around the vortex once, namely $\phi\to ve^{iN\theta}$, where $\theta$ is the coordinate polar angle. Thus, vortices are characterized by the integer $N$, called vorticity, which must be an integer if $\phi$ is to be continuous. These solutions are said to be topological because their asymptotic behavior cannot be deformed continuously to the vacuum solution $\phi=v$ and they are therefore stable. In fact, the vorticity is the degree of a map between the $S_{\infty}^1$ at spatial infinity and the $S_{\phi}^1$ of the complex phase. Following with the requirement of finiteness of \eqref{eq:energy_functional}, we conclude that the gauge field must behave asymptotically as $eA_i\to -N\partial_i \theta$ for the first two terms to vanish at infinity. This behavior implies that Abrikosov vortices inevitably carry a quantized magnetic flux
\begin{equation}
	\Phi = e\int d^2x\, B = -e\oint A_\theta d\theta = 2\pi N\;.
\end{equation}

To get the full solutions one must in general solve the field equations, which form a system of coupled second-order differential equations, imposing the vortex asymptotic boundary conditions. However, in the particular case where the coupling constants satisfy
\begin{equation}
	\lambda=2e^2\;,
\end{equation}
known as the Bogomol'nyi self-dual point \cite{bogomolnyi}, one can rewrite \eqref{eq:energy_functional} as
\begin{equation}
	\mathcal{E}=\int d^2x\left[\frac{1}{2}\left(B\pm e\big(\left|\phi\right|^2-v^2\big)\right)^2+\big|\left(D_1\pm iD_2\right)\phi\big|^2\pm ev^2B\right]\;.
	\label{eq:bogomol_trick}
\end{equation}
Then, minimizing the energy functional becomes trivial and yields a simpler system of first-order differential equations known as the Bogomol'nyi self-duality equations \cite{bogomolnyi},
\begin{align}
	\left(D_1\pm iD_2\right)\phi &= 0 \nonumber \\
	 e\left( \left| \phi\right|^2-v^2\right) &= \mp B  \;.
	\label{eq:bogomol}
\end{align}
The energy of the vortex (or its mass, since it is static) in this case is easy to compute and it is proportional to its magnetic flux, 
\begin{equation}
	M=\pm v^2 \Phi=\pm 2\pi Nv^2\;,
	\label{eq:flatmass}
\end{equation}
where the sign is to be chosen depending on $N$ to get a positive energy.

At this point, we would like to stress the similarities between Abrikosov vortices and higher-dimensional 't\ Hooft-Polyakov monopoles \cite{hooft,polyakov}. These appear, for example, in a (3+1)-dimensional $SU(2)$ gauge theory with an adjoint scalar field and a symmetry-breaking potential,
\begin{equation}
	S=\int d^4x \left[-\frac{1}{4}F_{\mu\nu}^aF^{a\,\mu\nu} + \frac{1}{2}D_\mu \phi^a D^\mu \phi^a - \frac{\lambda}{4}\left(\phi^a\phi^a-v^2\right)^2\right]\;.
	\label{eq:SU(2)_theory}
\end{equation}
Clearly, this action is analogous to \eqref{eq:flataction}, but with a Yang-Mills term instead of the usual Maxwell term and the covariant derivative in the adjoint representation, $D_\mu\phi^a=\partial_\mu \phi^a -e \epsilon_{abc}A^b_\mu \phi^c$. Looking for finite-energy solutions implies again that they live in the symmetry-broken phase, $\phi^a\phi^a\to v^2$, but here this leaves the freedom of a full sphere $S_{\phi}^2$ in field space. Just like with vortices, monopole solutions arise when the configuration of the scalar field at spatial infinity is topologically nontrivial and they are thus labelled by the degree of the map $S_{\infty}^2\to S_{\phi}^2$. Solving for the asymptotic behavior of the gauge field, one finds that these solutions carry a radial magnetic field of an unbroken $U(1)$ symmetry, which is precisely what awards them the name ``monopole'', and they carry a quantized magnetic flux but no electric charge. So the resemblance between the two solitons is obvious and it is therefore reasonable to expect that, just like self-gravitating 'tHP monopoles, self-gravitating Abrikosov vortices can hold a small BH at their core.

In this paper, we aim to generalize \eqref{eq:flataction} to curved space and study the effects of gravity on Abrikosov vortices following the classical analyses for 'tHP monopoles \cite{lee,ortiz,forgacs}. To that end, we consider the Abelian-Higgs model minimally coupled to gravity and turn on a Hilbert-Einstein term to make the metric dynamical. In addition, we will be interested in asymptotically AdS space-times, so we allow for a negative cosmological constant $\Lambda=-\frac{1}{\ell^2}$, where $\ell$ will denote the radius of the asymptotic AdS space-time.
\begin{align}
	S=\frac{1}{16\pi G}&\int d^3x \sqrt{g}\left(R+\frac{2}{\ell^2}\right) +\nonumber \\
	+ &\int d^3x \sqrt{g}\left[-\frac{1}{4}g^{\mu\rho}g^{\nu\sigma}F_{\mu\nu}F_{\rho\sigma} + g^{\mu\nu}D_\mu \phi\left(D_\nu \phi\right)^* - \frac{\lambda}{4}\left(|\phi|^2-v^2\right)^2\right]\;.
	\label{eq:curvedaction}
\end{align}
We use the signature commonly used in vortex literature $\left(+,-,-\right)$ and our conventions for the scalar curvature $R$ are as follows:
$$\Gamma\indices{_{\mu\sigma}^\nu}=\frac{1}{2}g^{\nu\tau}\left(\partial_\mu g_{\sigma\tau} + \partial_\sigma g_{\mu\tau} -\partial_\tau g_{\mu\sigma}\right)\,,$$
\begin{equation}
	R\indices{_{\mu\nu\rho}^\sigma} = \partial_\mu \Gamma\indices{_{\nu\rho}^\sigma} + \Gamma\indices{_{\mu \tau}^\sigma} \Gamma\indices{_{\nu\rho}^\tau} -\left(\mu \leftrightarrow \nu\right) \,,
	\qquad 
	R_{\mu\rho}= R\indices{_{\mu\sigma\rho}^\sigma}\,,
	\qquad R=g^{\mu\nu}R_{\mu\nu}\;.
\end{equation}

Our goal is to get the full vortex solutions of this theory by solving its field equations. But that is too complicated in the general case, so we will assume radial symmetry and work in polar coordinates $(t,r,\theta)$. For the matter fields, we take an ansatz that captures the features of a vortex solution, namely its vorticity and that it does not carry electric charge,
\begin{subequations}
\begin{align}
	\phi(t,r,\theta) \equiv & \;\sqrt{\rho(r)}\;e^{iN\theta} \;,\\
	e A_\mu(t,r,\theta)dx^\mu \equiv & \;\left(a(r)-N\right)d\theta \;.
\end{align}
\label{eq:matteransatz}
\end{subequations}
Note that even though the terms in $N$ look pure-gauge, they cannot be fully removed by non-singular gauge transformations, as the polar angle $\theta$ is not well defined at the origin. For the metric, we consider a Schwarzschild-like radial ansatz
\begin{equation}
	ds^2 = h(r)^2 dt^2 - \frac{1}{f(r)^2}dr^2 - r^2d\theta^2\;,
	\label{eq:metricansatz}
\end{equation}
where $g_{\theta\theta}=r^2$ corresponds to a coordinate ``gauge choice'' that defines the radial coordinate $r$ through the perimeter of full circles in $\theta$.

The next step would be to substitute \eqref{eq:matteransatz} and \eqref{eq:metricansatz} into the general field equations, but it is easier to plug them directly into the action \eqref{eq:curvedaction}, leaving
\begin{align}
	S=\int dtdrd\theta &\left[\frac{1}{8\pi G}fh'+\frac{1}{8\pi G}\frac{1}{\ell^2}\frac{rh}{f}-\right.\nonumber\\ &\;\left.-\frac{1}{2e^2}\frac{fh}{r}\left(a'\right)^2 -\frac{1}{4}rfh \frac{\left(\rho'\right)^2}{\rho} -\frac{h}{rf}a^2\rho- \frac{\lambda}{4}\frac{rh}{f}\left(\rho-v^2\right)^2\right]\;,
	\label{eq:curvedaction2}
\end{align}
(where we have dropped a boundary term), and derive the four coupled field equations by taking variations with respect to each of the fields in it. One can check that both procedures yield the same system of equations. Before writing it out, though, it is useful for the imminent numerical analysis to group the parameters of the theory into dimensionless ratios. We define an ``effective vortex radius'' and dimensionless coupling constants by
\begin{equation}
	r_0\equiv \frac{1}{\sqrt{2e^2v^2}}\,, \qquad \chi\equiv 16\pi G v^2\,, \qquad \gamma\equiv \frac{\lambda}{2e^2}\;,
	\label{eq:params}
\end{equation}
and we then make the rescalings $r\to r_0\,r$, $\ell\to r_0\,\ell$ and $\rho\to v^2\,\rho$.

Combining the field equations for the metric components, we obtain
\begin{equation}
\partial_r\left(\frac{h}{f}\right)= \chi \left( \frac{\left(a'\right)^2}{r}+\frac{r}{4}\frac{\left(\rho'\right)^2}{\rho}\right)\frac{h}{f}\;,
\label{eq:h_f_eq}
\end{equation}
which can be used to eliminate $h$ from all the remaining equations. The system of equations then reduces to
\begin{subequations}
\begin{equation}
	\partial_r\left(f^2\frac{r}{2}\frac{\rho'}{\rho}\right) = -\chi \left( \frac{\left(a'\right)^2}{r}+\frac{r}{4}\frac{\left(\rho'\right)^2}{\rho}\right)f^2\frac{r}{2}\frac{\rho'}{\rho} -f^2\frac{r}{4}\frac{(\rho')^2}{\rho^2}+\frac{a^2}{r}+\frac{\gamma}{2}r\left(\rho-1\right)\;,
\end{equation}
\begin{equation}
\partial_r\left(f^2\frac{a'}{r}\right) = -\chi\left(\frac{\left(a'\right)^2}{r}+\frac{r}{4}\frac{\left(\rho'\right)^2}{\rho}\right)f^2\frac{a'}{r} +\frac{a}{r}\rho\;,
\end{equation}
\begin{equation}
	\partial_r\left(f^2\right) = 2\frac{r}{\ell^2} -\chi \left(f^2\frac{\left(a'\right)^2}{r} +f^2\frac{r}{4} \frac{\left(\rho'\right)^2}{\rho} +\frac{a^2}{r}\rho+ \frac{\gamma}{4}r\left(\rho-1\right)^2\right)\;.
	\label{eq:f2_field_eq}
\end{equation}
\label{eq:field_equs}
\end{subequations}
We just need to solve this system for $\rho(r)$, $a(r)$ and $f^2(r)$ and then calculate $h^2(r)$ using
\begin{equation}
	h^2(r) = C f^2(r)\exp\left[2\int_0^r dr' \chi\left(\frac{\left(a'\right)^2}{r'}+\frac{r'}{4}\frac{\left(\rho'\right)^2}{\rho}\right)\right]\;,
	\label{eq:h2_eq}
\end{equation}
where the integration constant depends on the boundary conditions, $C=h^2(0)/f^2(0)$, but can be removed by rescalings of $t$.

The system of equations \eqref{eq:field_equs} is composed of one first-order and two second-order differential equations, so we must supply one boundary condition for the inverse metric component $f^2$ and two for each of the matter fields $\rho$, $a$. We will discuss these conditions in the coming sections for the different limits that the parameters in \eqref{eq:field_equs} allow. In particular, we will first consider asymptotically flat space-times by taking $\ell\to\infty$ (Section~\ref{sec:flat_section}) and then move to asymptotically AdS space-times with $\ell<\infty$ (Section~\ref{sec:ads_section}). In either case, we can decouple the metric and study the theory in a fixed background by taking $\chi=0$. For later reference, we list here the set of first derivatives needed to solve \eqref{eq:field_equs} numerically with an iterative Runge-Kutta method \cite{RK} once initial conditions for each field have been specified:
\begin{align}
	\partial_r \rho =&\, \rho' \nonumber\\
	\partial_r a =&\, a'      \nonumber\\
	\partial_r\rho' =&\, -\frac{\rho'}{r}+\frac{1}{2}\frac{(\rho')^2}{\rho}  + \frac{1}{f^2}\left[2\frac{a^2}{r^2}\rho+\gamma\rho\left(\rho-1\right) + \chi\left(\frac{a^2}{r}\rho \rho' + \frac{\gamma}{4}r\left(\rho-1\right)^2\rho'\right)-2\frac{r}{\ell^2}\rho'\right] \nonumber \\ 
	\partial_r a' =&\, \frac{a'}{r}+ \frac{1}{f^2}\left[ a\rho +\chi\left(\frac{a^2}{r}\rho a' + \frac{\gamma}{4}r\left(\rho-1\right)^2 a'\right)-2\frac{r}{\ell^2}a'\right]\nonumber \\
	\partial_r f^2 =&\, 2\frac{r}{\ell^2} -\chi \left[f^2\left(\frac{\left(a'\right)^2}{r} +\frac{r}{4} \frac{\left(\rho'\right)^2}{\rho}\right) +\frac{a^2}{r}\rho+ \frac{\gamma}{4}r\left(\rho-1\right)^2\right]\;.
	\label{eq:RK_field_equs}
\end{align}

\section{Asymptotically flat space-times} \label{sec:flat_section}
\subsection{Vortices in flat space-time} \label{sec:fixed_flat}
Let us start with the well-studied Abrikosov vortex in flat space to introduce the methods that we will use to solve \eqref{eq:field_equs} in the general case and for comparison with later results. The flat-space limit of \eqref{eq:field_equs} is obtained by taking $\chi=0$, $\ell\to\infty$ and $f^2(r)=1$, and only four boundary conditions are needed in this case. As discussed at the beginning of Section~\ref{sec:theory}, vortex solutions are characterized by the asymptotic behavior $\phi\to ve^{iN\theta}$, $eA_i\to -N\partial_i\theta$, which in terms of the ansatz \eqref{eq:matteransatz} implies the asymptotic boundary conditions
\begin{equation}
	\rho\left(r\to\infty\right)= 1\,,\qquad a\left(r\to\infty\right)= 0\;.
	\label{eq:asymptotic_bc}
\end{equation}
In addition, requiring the solutions to be regular fixes the two remaining boundary conditions,
\begin{equation}
	\rho\left(r=0\right)=0\,,\qquad a\left(r=0\right)=N\;,
	\label{eq:flat_origin_condition}
\end{equation}
since the polar angle $\theta$ is not well-defined at the origin. No exact solution is known even for the self-duality equations \eqref{eq:bogomol} at the Bogomol'nyi point $\gamma=1$, let alone for general $\gamma$ or when we later move on to curved space. But one can solve these equations numerically with iterative methods such as Runge-Kutta (RK) methods \cite{RK}.

\begin{figure}[tbp]
\centering
\includegraphics[scale=0.8]{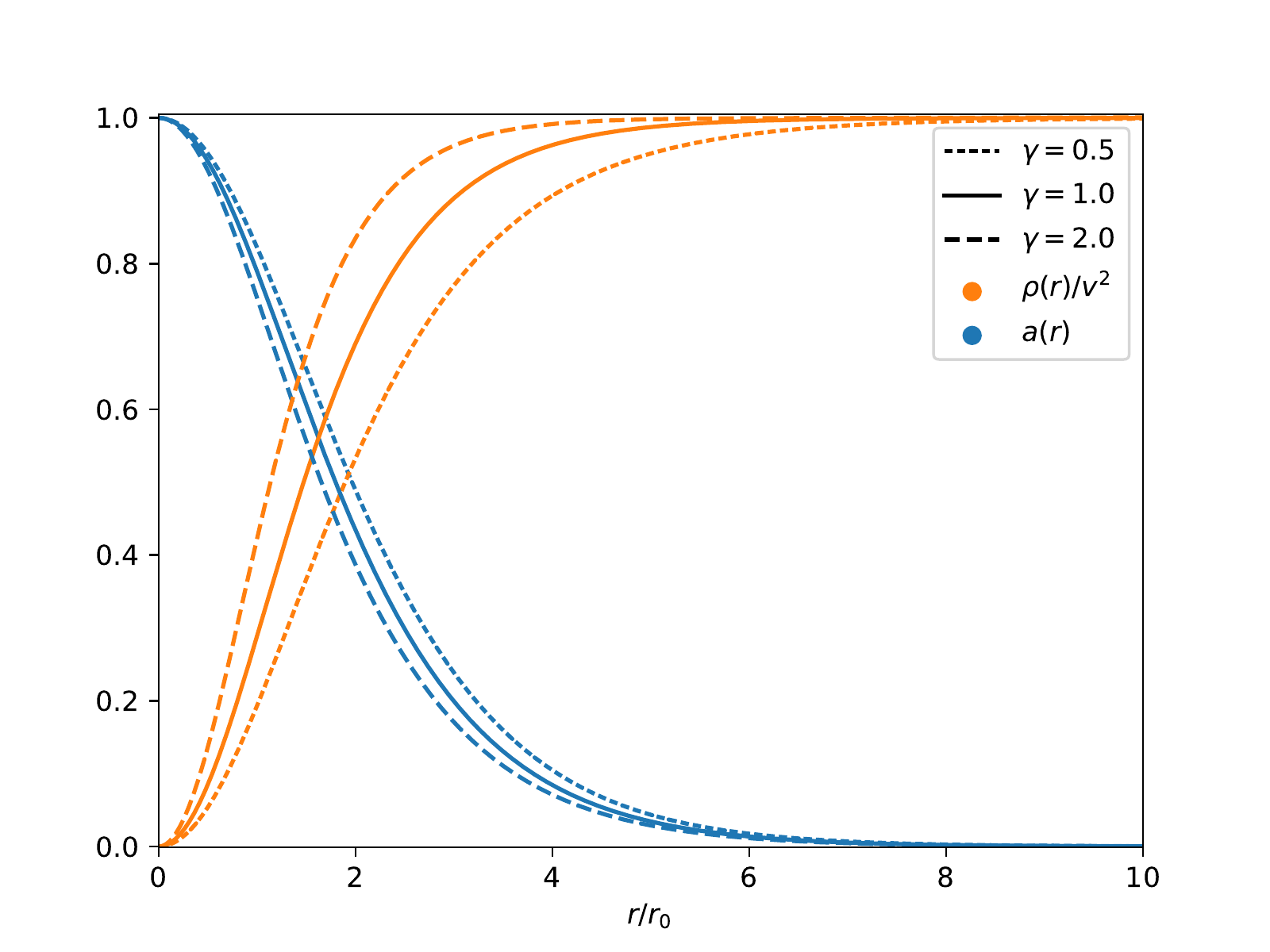}
\caption{\label{fig:flatprofile} Abrikosov vortex of vorticity $N=1$ for different values of the coupling ratio $\gamma=\lambda/(2e^2)$. Some parameters have been reintroduced for clarity.}
\end{figure}

However, RK methods need initial conditions (at $r=0$) for the functions and their derivatives, and our problem is a boundary value one. Expanding the field equations \eqref{eq:field_equs} in flat space around the origin, we obtain the small-$r$ behavior
\begin{equation}
	\rho(r)=C_\rho r^{\pm 2N} + \ldots\,, \qquad a(r)= N - C_a r^2 +\ldots \;,
\label{eq:flatexpansion}
\end{equation}
which can be used to take the first RK step given some values for the constants $C_\rho$, $C_a$. This way we can implement a ``shooting algorithm'' where we aim for the values of $C_\rho$, $C_a$ that make the solutions reach asymptotically \eqref{eq:asymptotic_bc}. In practice, one can assume that $\rho$ and $a$ are monotonic to do so, for we are interested in the solutions of lowest energy. We have implemented these procedures with an RK4 method of step $\Delta r = 0.0005\,r_0$; Figure~\ref{fig:flatprofile} shows our results for different values of the coupling ratio $\gamma$.

These solutions are in agreement with the extensive literature on Abrikosov vortices (see e.g.\ \cite{deVega}). In particular, changing $\gamma$ basically affects the fall-off of the scalar field $\rho$ toward the boundary. Indeed, taking the large-$r$ limit of \eqref{eq:field_equs} shows that the fields decay exponentially as
\begin{equation}
	\rho(r)-1\sim e^{-\sqrt{\gamma}\,r/r_0}\,,\qquad a(r)\sim e^{-r/r_0}\;.
	\label{eq:flat_decay}
\end{equation}
When restoring parameters, these decays can be explained in terms of the Higgs mechanism. Far away from the origin, the system is in the symmetry-broken phase and so the gauge field and the radial scalar mode acquire a mass $m_A^2=2e^2v^2$, $m_s^2=\lambda v^2$, respectively. Also at large $r$, the mass terms become dominant in the field equations and they result in the decay $\sim e^{-m r}$.

As an aside, we mention that the difference in the decays \eqref{eq:flat_decay} for $\gamma \lessgtr 1$ has been shown to determine the sign of the interaction between nearby vortices \cite{jacobs}. When $\gamma > 1$, close vortices repel each other while when $\gamma<1$, they attract, and in the context of superconductivity this relates to the distinction between Type I and Type II superconductors. In between these cases, the Bogomol'nyi self-dual point $\gamma=1$ is not only special in that it allows us to perform the ``Bogomol'nyi trick'' \eqref{eq:bogomol_trick} but also because the scalar and gauge fields decay in the same way at this point, allowing for static multivortex configurations. We note here that our solution for $\gamma=1$ satisfies the Bogomol'nyi self-duality equations \eqref{eq:bogomol}, as it should be. Having seen what the effect of changing $\gamma$ is, we will not study it further and we will henceforth stick to $\gamma=1$, unless otherwise specified.

\subsection{Asymptotically flat self-gravitating vortices} \label{sec:asymflat}
Keeping a vanishing cosmological constant, $\ell\to\infty$, we turn on $G$ to make the metric dynamical. In this case, to solve \eqref{eq:field_equs} we must supply, apart from four boundary conditions for the matter fields, an extra boundary condition for the metric component $f^2$. Asymptotically, the metric will be flat, so the same asymptotic boundary conditions for the matter fields \eqref{eq:asymptotic_bc} apply. But for the remaining boundary conditions, we can consider different possibilities. Following the studies of self-gravitating 'tHP monopoles \cite{lee,ortiz,forgacs}, we start with non-singular vortex solutions by choosing space-time to be flat at the origin and using the same regularity conditions for the matter fields \eqref{eq:flat_origin_condition} as before. Summarizing, the boundary conditions in this case are
\begin{equation}
	f^2(r=0)=1\,,\quad\rho\left(r=0\right)=0\,,\quad a\left(r=0\right)=N\,,\quad\rho\left(r\to\infty\right)= 1\,,\quad a\left(r\to\infty\right)= 0\;.
	\label{eq:asymflat_bdy_cond}
\end{equation}
With this setup, we can study the effects of gravity on the flat-space solutions discussed above as we turn up the ``strength of gravity'' $\chi$. The field equations can be solved in the same way as before, namely with an RK method combined with a shooting algorithm. Using that $f^2\sim 1$ at the origin, one can show that the expansion \eqref{eq:flatexpansion} still holds, so we still can use it to make the first step of the iterative method.

\begin{figure}[tbp]
\centering
\includegraphics[scale=0.8]{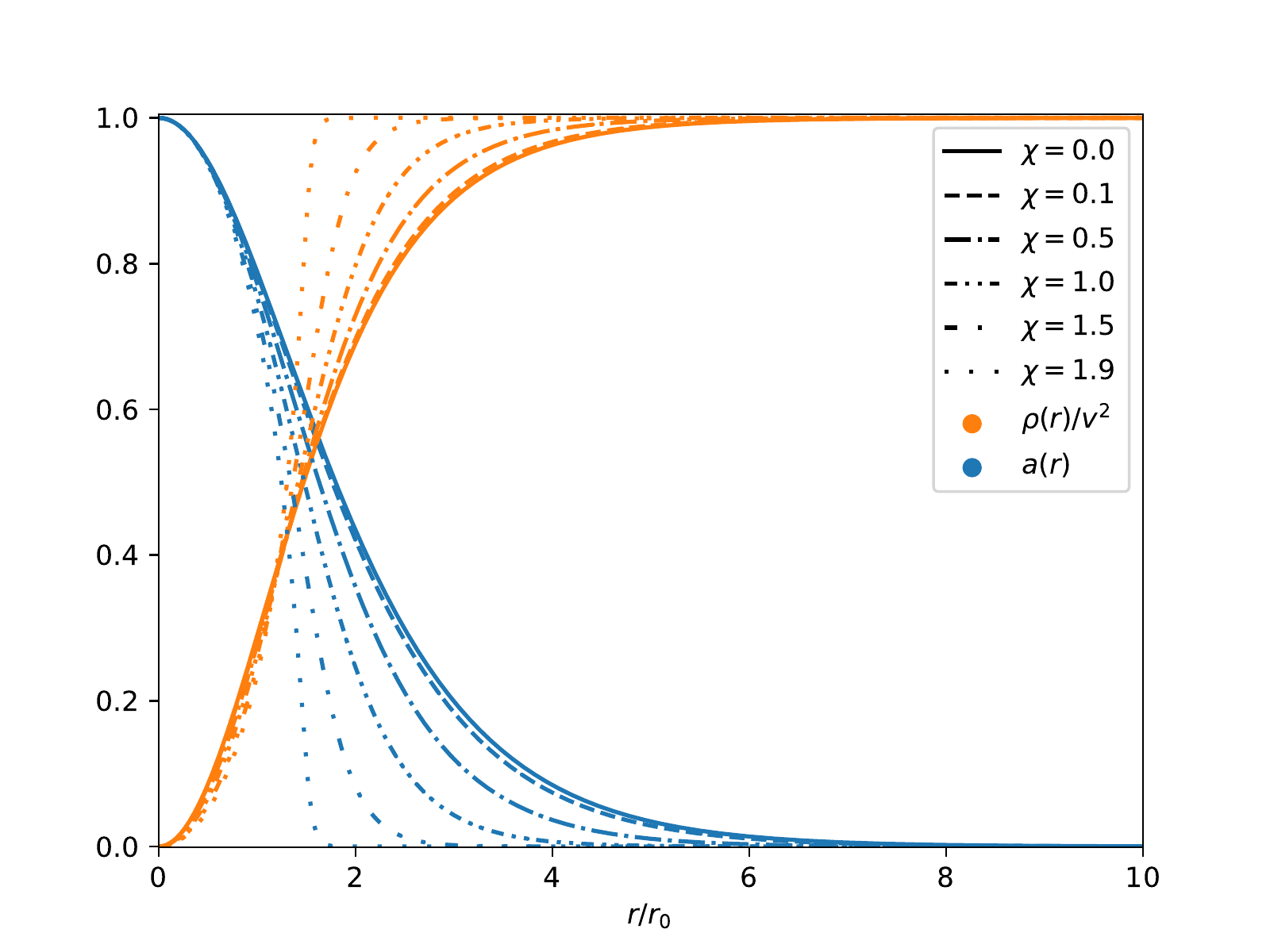}
\caption{\label{fig:asymflat_matter} Matter fields of the non-singular self-gravitating Abrikosov vortex of vorticity $N=1$ for increasing values of the ``strength of gravity'' $\chi=16\pi G v^2$.}
\end{figure}

\begin{figure}[tbp]
\centering
\includegraphics[scale=0.8]{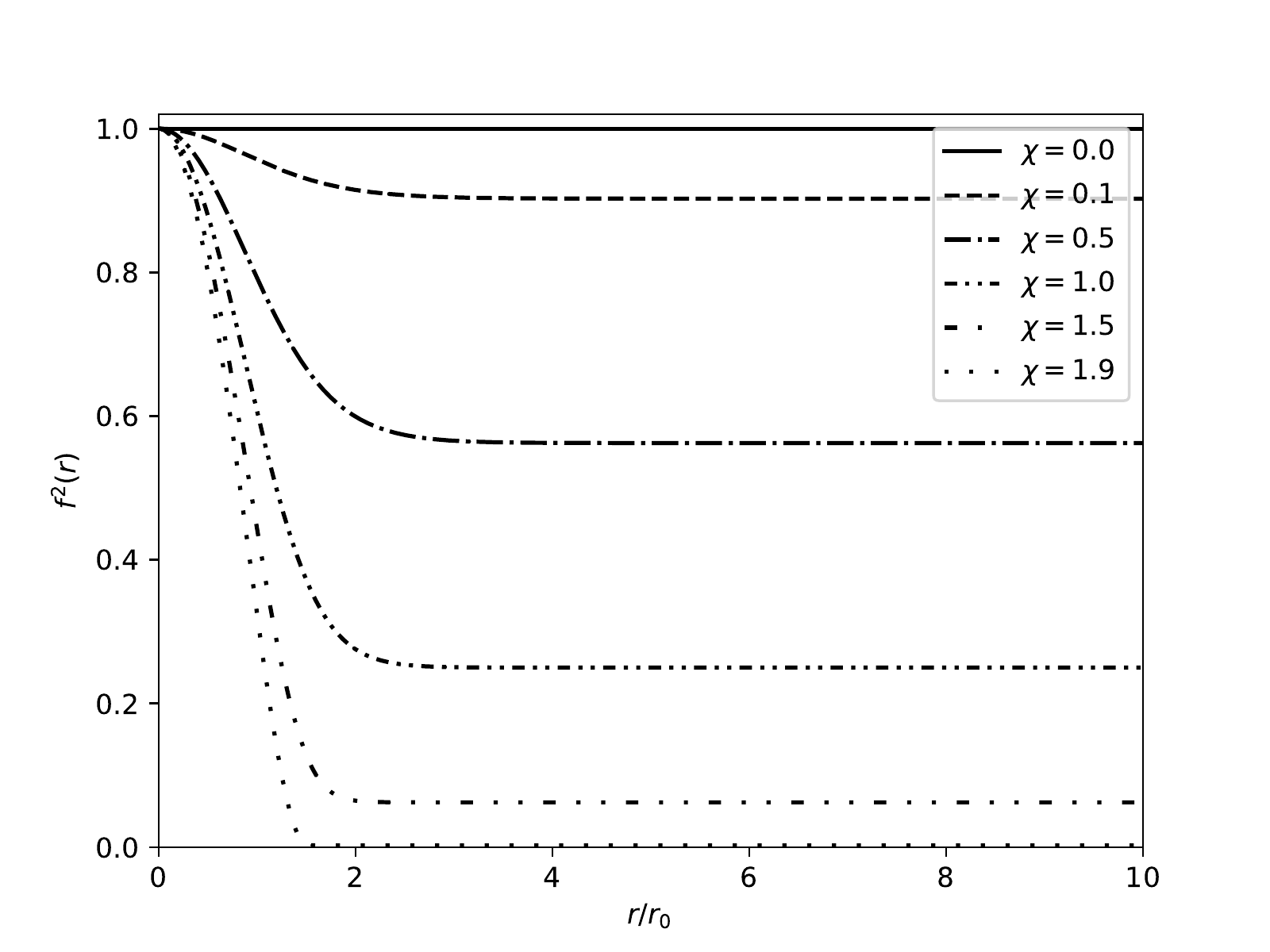}
\caption{\label{fig:asymflat_metric} Metric component of the non-singular self-gravitating Abrikosov vortex of vorticity $N=1$ for increasing values of the ``strength of gravity'' $\chi=16\pi G v^2$.}
\end{figure}

Figures \ref{fig:asymflat_matter} and \ref{fig:asymflat_metric} show our results for increasing values of $\chi$, depicting how an initially flat Abrikosov vortex changes as its mass increases, as well as its backreaction on the metric. With these solutions, a computation of $h^2(r)$ using \eqref{eq:h2_eq} and the boundary condition $h^2(r=0)=1$ yields the striking result $h^2(r)=1$ for every $\chi$. This only happens at the special self-dual point $\gamma=1$ that we are considering and it can already be seen by trying to rewrite the curved-space action \eqref{eq:curvedaction} using a ``Bogomol'nyi trick'' as in \eqref{eq:bogomol_trick}, \cite{valtancoli}. Doing that involves integrating by parts, but in contrast to the flat-space case, one must now move a derivative across $h(r)$. As a result, the action can only be rewritten as a sum of squares when $h(r)$ is constant. Again, our solutions satisfy the curved-space self-duality equations analogous to \eqref{eq:bogomol} that one gets by minimizing that functional \cite{valtancoli}, as it should be.

As either Newton's constant $G$ or the vev $v^2$ are increased, the vortex gets pulled inwards, reducing the effective radii of the matter distributions. At the same time, the radial component of the inverse metric $f^2$ tends asymptotically to a constant that decreases with $\chi$. To find said constant, it is useful to compute the curved-space conserved energy of these vortices (or mass, since they are static) associated to the temporal Killing vector $\partial_t$,
\begin{equation}
	M = \int_{\Sigma_t} d^2x \sqrt{g}\, T\indices{^0_{\,0}} = \int dr d\theta \frac{r}{f} \frac{1}{h}\,T_{00}\;,
\end{equation}
where the energy-momentum tensor is
\begin{equation}
T_{\mu\nu}= F_{\mu\rho}F\indices{^\rho_\nu}+\left( D_\mu\phi\right)^* D_\nu\phi + \left(D_\nu\phi\right)^* D_\mu\phi+ g_{\mu\nu} \bigg(\frac{1}{4}F_{\alpha\beta}F^{\alpha\beta} -\left| D_\alpha\phi\right|^2+ V\left[\phi\right]\bigg)\;.
\label{eq:EM_tensor}
\end{equation}
With the field equations \eqref{eq:field_equs}, its temporal component can be put in the form
\begin{equation}
T_{00}= -h^2\frac{f}{r}\partial_r\left(\frac{1}{8\pi G}f\right) \;,
\end{equation}
so --using that $h=1$ for the vortex solutions-- we get the asymptotic value
\begin{equation}
	f(\infty) = 1-4 G M\;.
	\label{eq:f2_asym}
\end{equation}
At the same time, the mass of the self-dual vortex in curved space can be derived as in Section \ref{sec:theory} from the action rewritten à la Bogomol'nyi \cite{valtancoli} and it is in fact the same as in flat space \eqref{eq:flatmass}. When plugging it into \eqref{eq:f2_asym} and relabelling parameters as in \eqref{eq:params}, we get $f^2(\infty)=\left(1-\chi N/2\right)^2$, which agrees with the asymptotic values of Fig.~\ref{fig:asymflat_metric}. 

\begin{figure}[tbp]
    \centering
    \subfloat[$M<\frac{1}{4G}$]{{\includegraphics[width=0.25\textwidth]{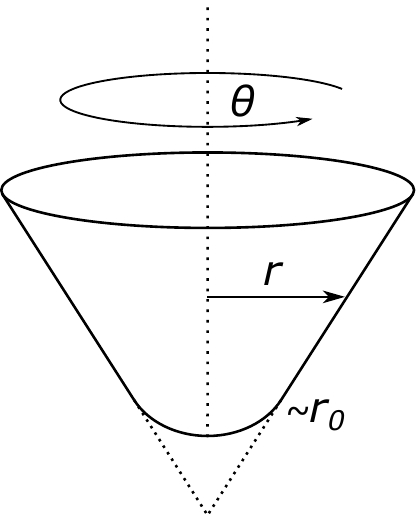}}} 
    \subfloat[$M=\frac{1}{4G}$]{{\includegraphics[width=0.25\textwidth]{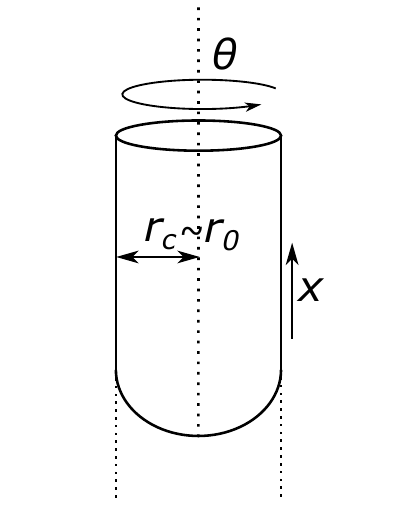}}}%
    \subfloat[$M>\frac{1}{4G}$]{{\includegraphics[width=0.25\textwidth]{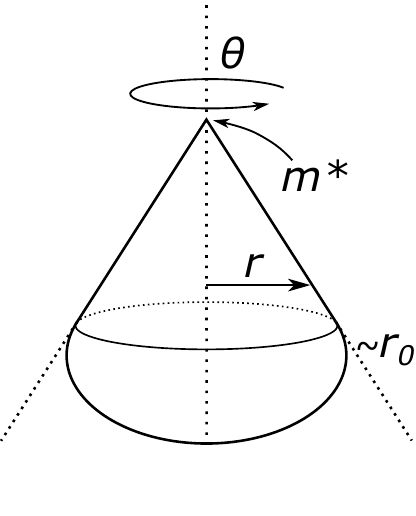}}}%
    \subfloat[]{{\includegraphics[width=0.25\textwidth]{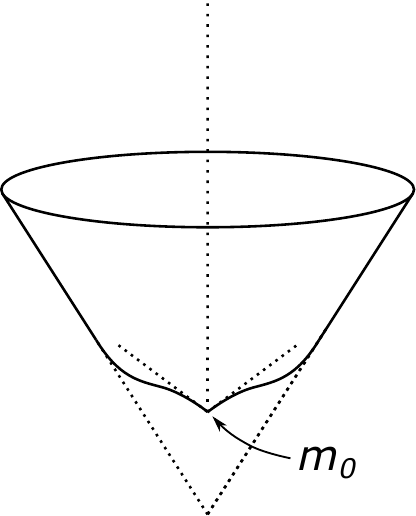}}}%
    \caption{Qualitative picture of the different spaces generated by self-gravitating Abrikosov vortices.}
    \label{fig:cones}
\end{figure}

Thus, the metric of the vortex is asymptotic to
\begin{equation}
	ds^2 = dt^2 - \frac{1}{\left(1-4GM\right)^2} dr^2 - r^2d\theta^2\;,
	\label{eq:cone_metric}
\end{equation}
which is the metric of a cone. Locally, this metric indeed corresponds to flat space\footnote{The coordinate transformation $\tilde{r}=\frac{r}{1-4GM}$, $\tilde{\theta}=\left(1-4GM\right)\theta$ brings the metric to $ds^2=dt^2-d\tilde{r}^2-\tilde{r}^2d\theta^2$, but the new angular coordinate only takes values in $\tilde{\theta}\in \left[0,\left(1-4GM\right)2\pi\right)$.}, but globally, it has a deficit angle of $8\pi GM$. This is nothing but the metric of a static point particle of mass $M$ in $2+1$ dimensions, as found by Deser and Cl\'{e}ment in the '80s \cite{deser,clement},
\begin{equation}
	ds^2=dt^2-\left|\vec{x}\right|^{-8GM}d\vec{x}^2\;.
	\label{eq:pointmass_metric}
\end{equation}
To see the equivalence, transform it to polar coordinates and redefine the radial coordinate $r\equiv \left|\vec{x}\right|^{(1-4GM)}$ to make it match with the gauge defined by the ansatz \eqref{eq:metricansatz}. At the origin, on the other hand, the non-singular vortex is truly flat. So the extended nature of vortex solutions replaces the conical singularity at the origin of the point particle's metric by a smooth drift to truly flat space, rounding off the tip of the cone (see Fig.~\ref{fig:cones} (a)).

In the 4-dimensional analogue of Fig.~\ref{fig:asymflat_metric} for 'tHP monopoles \cite{lee,ortiz,forgacs}, $f^2(r)$ has a minimum that decreases with $\chi$. As soon as the minimum reaches 0, one argues that the monopole has shrunk beyond its Schwarzschild radius and it forms a magnetic RN black hole with the exterior horizon at the point where $f^2(r_h)=0$. In our case of $D=2+1$ vortices, in contrast, $f^2(r)$ does not have a unique minimum and hence the story is rather different. As $GM$ increases, the asymptotic cone keeps closing until the critical value $M_{\text{crit}}=\frac{1}{4G}$ is reached ($\chi=2$ in Fig.~\ref{fig:asymflat_metric}). At this point, there is an asymptotic deficit angle of $2\pi$ and the space-time becomes a semi-infinite cylinder (see Fig.~\ref{fig:cones} (b)). The metric \eqref{eq:cone_metric} has a coordinate singularity in this limit. Indeed, making the transformation $x=\left(r-r_c\right)/\left(1-4GM\right)$ right before taking $GM\to 1/4$ yields the metric of a cylinder of radius $r_c$ \cite{peleg}, $ds^2=dt^2-dx^2-r_c^2d\theta^2$.

Beyond that point, the asymptotic side of the cylinder closes onto a cone that makes space-time compact (see Fig.~\ref{fig:cones} (c)). Despite the coordinate singularity $f^2=0$ that prevents us from applying the numerical analysis discussed above, it is clear that the solution in this case consists of an Abrikosov vortex inside a region $\sim r_0$ glued to an inverted cone that corresponds to a dual point mass with $m^*=\frac{1}{2G}-M$. Although fairly odd, this sort of space-times is indeed what one finds when considering extended masses such as dust shells in $D=2+1$ \cite{peleg}. It might look worrisome that these solutions occur for $v^2\gtrsim G^{-1}$, which seems to correspond to the Planck scale, where we expect physics to be modified by the quantum effects of gravity. But in $D=2+1$, the dimensions of $G$ are $M^{-1}L^2T^{-2}$ and the ``Planck mass'' ($m_p=c^2/G$) does not involve $\hbar$, so the semi-classical theory is still valid at these scales\footnote{We thank Roberto Emparan for pointing this out to us.}.

As a final remark, we note that changing $\gamma$ has no significant effects on the qualitative picture of Fig.~\ref{fig:cones}. As far as the matter fields are concerned, the decay of the scalar field with respect to the gauge field changes as in Fig.~\ref{fig:flatprofile} and, when turning up $\chi$, the situation is analogous to Fig.~\ref{fig:asymflat_matter}. As for the metric, $f^2$ shows a very similar behavior to that of Fig.~\ref{fig:asymflat_metric}, albeit with different asymptotic values and different critical $\chi$, as befits a different vortex mass. Meanwhile, $h^2$ ceases to be $1$ to reach asymptotically a constant value that is above (below) $1$ when $\gamma<1$ ($\gamma>1$). There is therefore some time dilation/contraction between the time at the center of the vortex and at infinity.

This is as far as non-singular solutions go, but we are most interested in singular solutions since, after all, we want to find out whether Abrikosov vortices can hold a BH at their core. Following \cite{lee,forgacs}, we consider the insertion of a small point mass at the core of the vortex by changing the boundary condition $f^2(0)$. As discussed in the Introduction, in $D=3+1$ this brings in a small Schwarzschild BH at the core of the 'tHP monopole and one can then study the stability of hair outside the horizon. But in $D=2+1$, to insert a point particle of mass $m_0$ at the origin we must take the boundary condition $f^2(0)=(1-4Gm_0)^2$, and we see from \eqref{eq:cone_metric} that this just brings back a conical singularity at the core of the vortex (see Fig.~\ref{fig:cones} (d)). It does not yield a BH at all. In a conical singularity the curvature does not blow up, so we still ought to demand regularity of the matter fields at the origin. Therefore, vortex solutions holding a conical singularity at their core are very similar to their non-singular counterparts.

At the end of the day, we see that asymptotically flat Abrikosov vortices cannot hold a Schwarzschild BH at their core for the simple reason that the Schwarzschild BH does not exist in 3-dimensional gravity. To completely rule out BH solutions, let us imagine that, somehow, we have managed to insert a BH inside an asymptotically flat Abrikosov vortex. This would define an (exterior) horizon $r_h$ where $f^2(r_h)=0$ and, from \eqref{eq:f2_field_eq},
\begin{equation}
	\left(f^2\right)'_h\equiv\left.\frac{d f^2}{dr}\right|_{r_h}= -\chi \left.\left(\frac{a^2}{r}\rho+ \frac{\gamma}{4}r\left(\rho-1\right)^2\right)\right|_{r_h}\;.
	\label{eq:f2p_asymflat_horizon}
\end{equation}
Asymptotically, we need $f^2>0$ for the radial coordinate $r$ to be spatial, so it better be that $\left(f^2\right)'_h\geq 0$. But \eqref{eq:f2p_asymflat_horizon} shows that this is only possible when the matter fields take their asymptotic values already at the horizon and, in this case, the full solution is trivial with $f^2(r)$ equal to 0 all the way to infinity. We conclude that no BH solution exists in this theory, so it does not really make sense to ask whether these vortices can realize BH hair.

\section{Asymptotically AdS space-times} \label{sec:ads_section}
The reason why no BH is found in asymptotically flat $D=2+1$ space-times is the peculiar dimensions of $G$. There is no possible way of constructing a length scale out of a mass and $G$ (and $c$). But BH solutions intrinsically need a length scale to define the horizon, so it is clear that ordinary BHs will not exist in $D=2+1$. Indeed, \eqref{eq:cone_metric} is the would-be Schwarzschild metric, but the combination $GM$ is dimensionless and hence it is not divided by $r$. A way to go around this issue and get a BH is to introduce a length scale by means of a cosmological constant. In particular, introducing a negative cosmological constant allows for an asymptotically AdS$_3$ BH solution first found by Bañados, Teitelboim and Zanelli; the BTZ black hole \cite{btz}. The next logical step is thus to add such a cosmological constant to our theory and study self-gravitating Abrikosov vortices in asymptotically AdS space-times to see how they relate to BTZ black holes. However, we will first consider vortices in fixed AdS$_3$ to discuss their boundary conditions and later we will make the metric dynamical.

\subsection{Vortices in fixed AdS\texorpdfstring{$_3$}{3}} \label{sec:fixed_ads}
Anti de Sitter (AdS) space-time is a manifold of constant negative curvature that solves Einstein's vacuum field equations with a negative cosmological constant $\Lambda=-\frac{1}{\ell^2}$, where the parameter $\ell$ is referred to as the AdS radius. To match with the ansatz \eqref{eq:metricansatz}, we consider global AdS$_3$ with metric
\begin{equation}
	ds^2 = \left(1+\frac{r^2}{l^2}\right)dt^2 - \frac{1}{\left(1+\frac{r^2}{l^2}\right)}dr^2 -r^2 \,d\theta^2\;.
	\label{eq:adsmetric}
\end{equation}
This way, we can read off $h^2$ and $f^2$ and use them --together with $\chi=0$ to fix the metric-- in the general field equations \eqref{eq:field_equs} to get the equations that describe Abrikosov vortices in fixed AdS$_3$. Just like in Section \ref{sec:fixed_flat}, where the metric was held fixed, we must supply four boundary conditions for the matter fields to specify a solution. As usual, two of them will come from the origin and the other two will come from the boundary. Given that AdS space-time \eqref{eq:adsmetric} is flat near the origin, we can impose the same regularity conditions \eqref{eq:flat_origin_condition} as before, but the asymptotic boundary conditions require a bit more discussion.

The main issue with AdS space-time is that it is not globally hyperbolic so, in general, matter fields propagating in it do not pose a well-defined Cauchy problem. The physical interpretation of this is that information can leak in/out from the boundary because the AdS metric allows radial light-rays to reach the boundary in a finite time\footnote{This is most easily seen in the coordinates $r=\ell \tan\rho$, for which the boundary sits at $\rho=\frac{\pi}{2}$.}. One must therefore be careful and impose boundary conditions to prevent this. The usual resolution is to require the energy functional
\begin{equation}
	E = \int_{\Sigma_t} d^2x \sqrt{g}\, T\indices{^0_{\,0}}
	\label{eq:curved_energy}
\end{equation}
to be conserved, positive and finite. In addition, one can define an inner product between mode solutions that should also be conserved \cite{breitenlohner}, which yields the notion of normalizable and non-normalizable modes; an important concept for the interpretation of bulk solutions in the context of AdS/CFT \cite{balasubramanian}. We will start with the requirements on \eqref{eq:curved_energy} as they will be enough to motivate the boundary conditions for vortices in AdS$_3$, and we postpone the more careful AdS/CFT analysis until after we have found the solutions (see Section \ref{sec:ads/cft}).

The integrand in \eqref{eq:curved_energy} is formally conserved\footnote{It is the 0th component of $j^\mu=\sqrt{g}T\indices{^\mu_ \nu}\xi^\nu$ (with $\xi^\nu=\delta^\nu_0$ a time-like Killing vector of AdS space-time), which is conserved ($\partial_\mu j^\mu=0$) by the covariant conservation of the energy momentum tensor and the Killing equation.}, so for $E$ to be conserved one must demand that its flux through the boundary vanishes. However, our Abrikosov vortices are static solutions by assumption, so their energy is trivially conserved in time and, recalling that these vortices have no electric fields, one can check that the boundary energy flux vanishes trivially as well. When plugging in the energy momentum tensor \eqref{eq:EM_tensor}, we get a covariantized version of \eqref{eq:energy_functional} that is again trivially positive. Thus, the only relevant requirement is finiteness of the energy and it actually implies the same vortex boundary conditions as in flat space, namely
\begin{equation}
	\phi\to ve^{iN\theta}\,\qquad eA_i\to -N\partial_i \theta\;.
	\label{eq:ads_bdy_cond}
\end{equation}

\begin{figure}[tbp]
\centering
\includegraphics[scale=0.8]{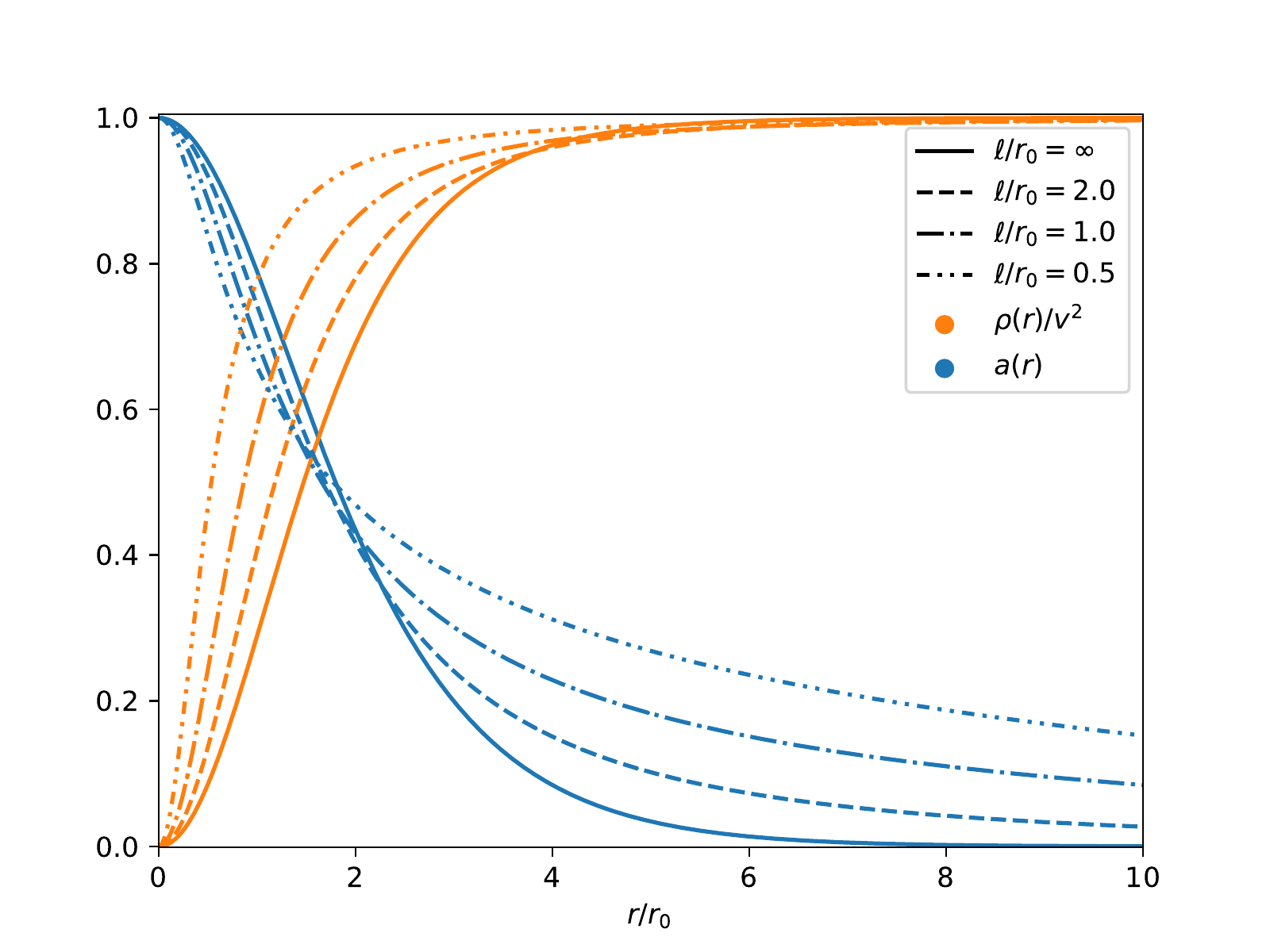}
\caption{\label{fig:adsprofile} Abrikosov vortex of vorticity $N=1$ in fixed AdS$_3$ for different values of the AdS radius $\ell$.}
\end{figure}

In sum, the boundary conditions for Abrikosov vortices in fixed AdS$_3$ are as in Section \ref{sec:fixed_flat} and so an analogous numerical analysis holds. In particular, the first RK step can be taken in the same way because the small-$r$ behavior from flat space \eqref{eq:flatexpansion} survives in AdS$_3$. In Fig.~\ref{fig:adsprofile}, we plot our solutions for different values of the AdS radius $\ell$. The main effect that AdS space-time has on the Abrikosov vortex is to change the decay of its fields toward their asymptotic values. As the AdS radius is reduced (i.e.\ as the curvature increases), the scalar field gets pulled inwards while the gauge field expands to cover a larger region. The physical reason for this is that vortices owe their existence to a balance between the attractive pressure of the scalar field and the repulsive pressure of the gauge field \cite{gibbons}. When put in AdS space-time, they feel an additional gravitational attractive pressure that shifts the balance in favor of a larger distribution for the gauge field.

In this case, the large-$r$ limit of \eqref{eq:field_equs} shows that the fields decay in a power law, in contrast to the exponential decay \eqref{eq:flat_decay} of flat space,
\begin{equation}
	\rho(r)-1\sim r^{-1\mp \sqrt{1+\gamma \ell^2/r_0^2}}\,,\qquad a(r)\sim r^{-\ell/r_0}\;.
	\label{eq:ads_decay}
\end{equation}
As in flat space, though, these decays are due to the masses that the fields get via the Higgs mechanism toward the boundary. This mechanism works in the same way as before and it gives the fields the same masses $m_A^2=2e^2v^2$, $m_s^2=\lambda v^2$, but a massive field in AdS decays in a power law rather than exponentially. From \eqref{eq:ads_decay}, we see that the main effect of $\gamma$ is again on the decay of the scalar field, but we note that $\gamma=1$ is no longer special. The fields no longer decay in the same way and the action cannot be rewritten with a ``Bogomol'nyi trick'' anymore because one has to move a derivative across $h(r)$ that does not vanish now. Perhaps there exists some modification of the scalar's potential (that becomes trivial in flat space) that can cancel this new term and make Abrikosov vortices remain self-dual in AdS space-time, but we have not explored this possibility.

When $\gamma>0$, only the upper sign for the decay of $\rho$ in \eqref{eq:ads_decay} is allowed by the asymptotic boundary condition $\rho\to 1$, whereas when $-r_0^2/\ell^2<\gamma<0$, either sign seems possible and we should expect different vortex solutions for each of these decays. In flat space tachyons are unstable, hence flat vortex solutions are only allowed for $\lambda>0$ (or $\gamma>0$). Otherwise, they would be unstable and the system would return to the symmetric phase. In AdS space-time, in contrast, tachyons can be stable so long as their mass is above the Breitenlohner-Freedman (BF) bound $m^2\ell^2>-1$ \cite{breitenlohner}. Therefore, AdS$_3$ stable vortices should also be possible in the window $-1<\lambda v^2 \ell^2<0$, which corresponds precisely to the range where both signs in \eqref{eq:ads_decay} are allowed. It is not immediately clear, however, whether they would be truly stable or just metastable since the symmetric phase is also stable in this range \cite{carmi}. In any case, we will center our discussion on $\lambda>0$ throughout this work.

\subsection{Asymptotically AdS self-gravitating vortices} \label{sec:asym_ads}
\begin{figure}[tbp]
\centering
\includegraphics[scale=0.8]{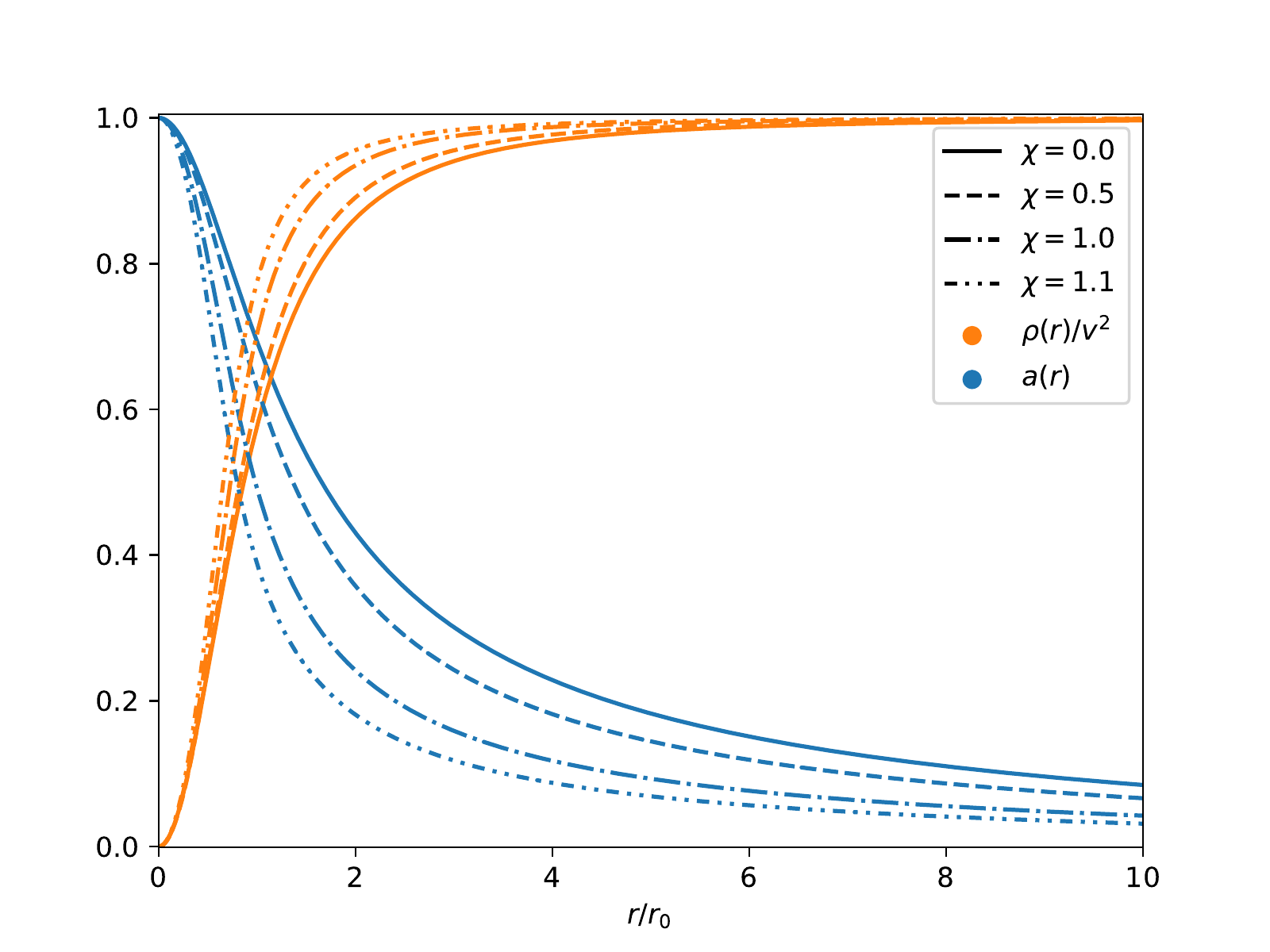}
\caption{\label{fig:asymads_matter} Matter fields of the non-singular self-gravitating Abrikosov vortex of vorticity $N=1$ in asymptotically AdS$_3$ space-time of radius $\ell=1$ for increasing values of the ``strength of gravity'' $\chi=16\pi G v^2$.}
\end{figure}

\begin{figure}[tbp]
\centering
\includegraphics[scale=0.8]{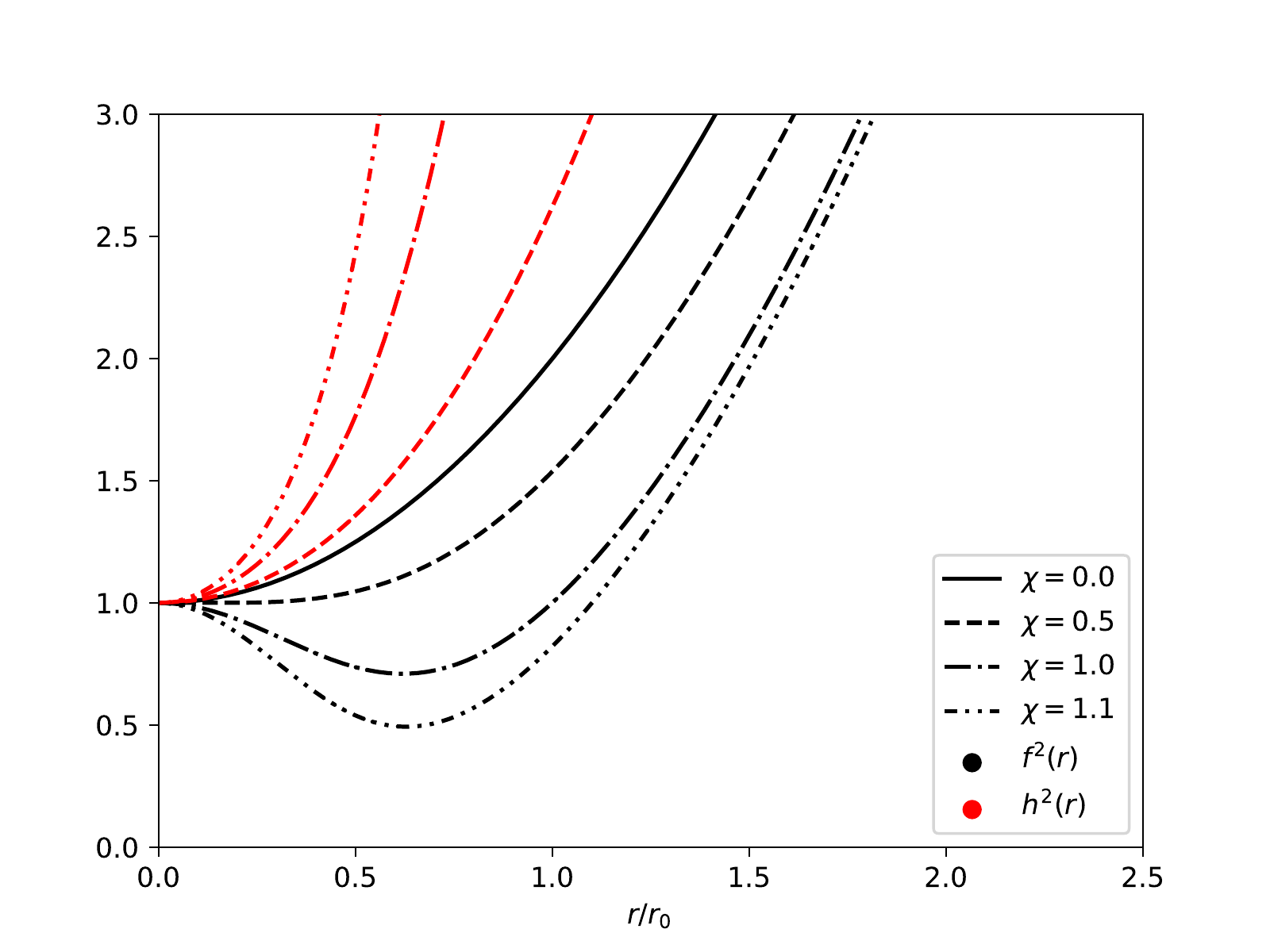}
\caption{\label{fig:asymads_metric} Metric components of the non-singular self-gravitating Abrikosov vortex of vorticity $N=1$ in asymptotically AdS$_3$ space-time of radius $\ell=1$ for increasing values of the ``strength of gravity'' $\chi=16\pi G v^2$.}
\end{figure}

Just like in Section \ref{sec:asymflat}, to make the metric dynamical, we must turn on $G$ and supply a boundary condition for the metric; but we now keep $\ell$ finite. Asymptotically, the matter fields will decay and cease to contribute to \eqref{eq:f2_field_eq}, leaving the quadratic behavior $f^2\to r^2/\ell^2$ that characterizes AdS space-time \eqref{eq:adsmetric}. Therefore, the same asymptotic vortex boundary conditions \eqref{eq:ads_bdy_cond} from the previous section apply, but there are different possibilities for the three remaining boundary conditions. Once more, we start with non-singular solutions by choosing space-time to be flat at the origin (i.e.\ $f^2(0)=1$) and fixing the regularity conditions \eqref{eq:flat_origin_condition} that we introduced in Section \ref{sec:fixed_flat}. Thus, the problem is essentially the same as before and it can be solved analogously. The matter fields and the metric components of a sample solution are shown in Figs.~\ref{fig:asymads_matter} and \ref{fig:asymads_metric} respectively. Like in the asymptotically flat case, turning up the ``strength of gravity'' pulls the vortex inwards as well as initially lowers $f^2$. However, at one point in $r$, the cosmological constant takes over and $f^2$ ends up growing quadratically, as befits asymptotically AdS space-times.

The general spherically-symmetric static solution to the free Einstein's equations with a negative cosmological constant is \cite{peleg}
\begin{equation}
	ds^2 = \left(C+\frac{r^2}{l^2}\right)dt^2 - \frac{1}{\left(C+\frac{r^2}{l^2}\right)}dr^2 -r^2 \,d\theta^2\;,
	\label{eq:BTZmetric}
\end{equation}
where $C$ is a constant that depends on a source located around the origin. When $C=1$, this is the source-free AdS$_3$ space-time \eqref{eq:adsmetric}. When $0<C<1$, much like in \eqref{eq:cone_metric}, there is a conical singularity at the origin corresponding to a point particle of mass $m=\frac{1}{4G}\left(1-\sqrt{C}\right)$. And when $C<0$, this corresponds to a non-rotating BTZ black hole of ADM mass $M=-C$ \cite{btz}. From Fig.~\ref{fig:asymads_metric}, it is clear that non-singular vortex solutions tend asymptotically to the metric \eqref{eq:BTZmetric}, albeit with a rescaling of the time coordinate. But here it is not straightforward to relate $C$ to global properties of the vortex because it depends not only on the mass of the source but also on its size (see e.g.\ \cite{peleg} for the case of collapsing dust shells).

These non-singular solutions with increasing mass and decreasing size are not expected to exist for arbitrary high values of $\chi$. Indeed, we find numerically that there is a critical value of $\chi$ beyond which the solutions for any values of $C_{\rho},C_a$ turn over too fast to reach (at least monotonically) the asymptotic boundary conditions. This is roughly the last value of $\chi$ shown in Fig.~\ref{fig:asymads_matter}. When considering different values of $\ell$, the situation is essentially the same but with different critical values for $\chi$. Thus, we expect that this theory will be taken over by singular solutions at one point. We turn our attention to these now.

Similarly to Section \ref{sec:asymflat}, taking the boundary condition $f^2(0)=(1-4Gm_0)^2$ corresponds to inserting a point mass at the origin that induces a conical singularity there (recall Fig.~\ref{fig:cones} (d)). This allows for solutions very similar to the non-singular ones described above that also exist only up to a critical value of $\chi$. What is more interesting, though, is to consider a negative boundary condition $f^2(0)<0$. In contrast to the asymptotically flat case, here this has the natural interpretation of inserting a BTZ black hole of ADM mass $M\sim-f^2(0)$ inside the vortex. The reason is that by the asymptotic AdS quadratic growth of $f^2(r)$, taking $f^2(0)<0$ inevitably yields a horizon $r_h$ where $f^2(r_h)=0$ and $(f^2)'_h>0$, evading the argument \eqref{eq:f2p_asymflat_horizon} that ruled out BH solutions in asymptotically flat space-times. So now it does make sense to ask whether Abrikosov vortices can provide hair for the BTZ black hole.

Physically, we must demand that matter fields be regular at the horizon. But the field equations \eqref{eq:field_equs} only allow us to choose five boundary conditions. So, if we want to hold onto the asymptotic vortex behavior, we must give up on fixing the fields at $r=0$. The problem then becomes a boundary value one between infinity and the (exterior) horizon $r_h$, where we can study whether non-trivial configurations for the matter fields (a.k.a.\ ``hair'') are possible. In the '90s, it was found that in the case of 4-dimensional 'tHP monopoles such configurations do exist \cite{lee,forgacs}. As briefly mentioned in the Introduction, at large $r_h$ there is only the hairless magnetic RN solution, where the matter fields remain constant at their asymptotic values. But when $r_h$ is small enough, there exists another solution in which the matter fields embody the tail of the monopole ``leaking out'' of the horizon (see Fig.\ 8 in \cite{forgacs}). This realizes a magnetic BH with hair that, in fact, is energetically preferred over the hairless solution. Given that Abrikosov vortices are analogous to 'tHP monopoles, it seems reasonable to expect that the same will hold for them and that they will ``dress'' the BTZ black hole.

Following \cite{lee}, we now study under what circumstances we could get such solutions. Given some boundary conditions $\rho_h,\, a_h$ for the fields at the horizon $r_h$, requiring regularity fixes the derivatives there\footnote{The brackets that are multiplied by $1/f^2$ in the field equations \eqref{eq:RK_field_equs} must vanish at $r_h$.},
\begin{subequations}
\begin{align}
	\left(f^2\right)_h' = &\, 2\frac{r}{\ell^2} -\chi \left(\frac{a_h^2}{r_h}\rho_h+ \frac{\gamma}{4}r_h\left(\rho_h-1\right)^2\right) \\
	\rho_h' = &\, \frac{1}{\left(f^2\right)_h'}\left(2\frac{a_h^2}{r_h^2}\rho_h + \gamma\rho_h\left(\rho_h-1\right)\right) \\
	a_h' = &\, \frac{1}{\left(f^2\right)_h'}a_h\rho_h\;.
\end{align}
\end{subequations}
So we would have enough information to integrate \eqref{eq:RK_field_equs} radially outward from the horizon with an RK method. To meet the asymptotic boundary conditions, in this case, one would just have to tune $\rho_h,\, a_h$ with a shooting algorithm like we did before with $C_\rho$, $C_a$. However, we can place stringent bounds on their possible values already from general considerations. Assuming, like in the sections above, that $\rho(r)$ and $a(r)$ are monotonic (and respectively take values in the ranges $[0,1]$, $[0,N]$), we must demand $\rho_h'\geq 0$ and $\text{sgn} (a_h)a_h' \leq 0$ for them to be able to reach their asymptotic values. Then, recalling that $\left(f^2\right)_h'$ is always positive at the (exterior) horizon, we see that the second condition can only be satisfied when either $\rho_h$ or $a_h$ vanish.

When $\rho_h=0$, the derivatives of the matter fields at the horizon vanish. Then, expanding in $\varrho = r- r_h$ as
$$f^2\left(r_h+\varrho\right)\sim \left(f^2\right)_h'\varrho +\dots\,, \quad \rho\left(r_h+\varrho\right)\sim C^\rho_n \varrho^n + \dots\,,\quad a\left(r_h+\varrho\right)\sim a_h + C^a_m \varrho^m + \dots\;,$$
it is easy to show that the matter fields vanish order by order leaving $\rho(r)=0$, $a(r)=a_h$. This solution corresponds to a BTZ black hole in the symmetric phase, so it clearly does not satisfy the asymptotic vortex boundary conditions and it must be discarded. Only the case $a_h=0$ remains, for which the condition on $\rho_h'$ implies $\rho_h=1$. In this case, the derivatives of the matter fields vanish as well and the full solution is the trivial constant solution with the fields at their asymptotic values; $\rho(r)=1$, $a(r)=0$. Surprisingly, in sharp contrast with 'tHP monopoles, we have found that Abrikosov vortices do not provide black hole hair! Whenever a horizon is turned on, the solution collapses to its asymptotic values, leaving no non-trivial behavior of the matter fields outside the horizon.

One might complain that our derivation hinges on the assumption of monotonicity of the matter fields and that we can perhaps have non-monotonic vortex hair for the BTZ black hole. But that is not the case. Singular Abrikosov vortices obey a no-hair theorem. Recalling the classical derivation of these theorems \cite{bekenstein1}, we multiply the field equation of $a(r)$,
\begin{equation}
	\partial_r\left(\sqrt{g}\frac{f^2}{r^2}a'\right)=\sqrt{g}\,2e^2a\left|\phi\right|^2\;,
\end{equation}
by $a(r)$ and integrate it between the horizon $r_h$ and infinity. Rearranging the terms, we obtain
\begin{equation}
	\left.\sqrt{g}\frac{f^2}{r^2}aa'\right|_{r_h}^\infty=\int_{r_h}^\infty dr\sqrt{g}\left[\frac{f^2}{r^2}\left(a'\right)^2+2e^2a^2\left|\phi\right|^2\right]\;.
	\label{eq:nohair}
\end{equation}
The term in the left hand side of \eqref{eq:nohair} vanishes asymptotically for vortex solutions by the decay \eqref{eq:ads_decay} and at the horizon by $f^2(r_h)=0$ (and the regularity of the matter fields). So the right hand side must vanish as well. Being a sum of positive terms, this implies that each term has to vanish separately for all $r>r_h$, and the only solution in agreement with the asymptotic vortex conditions is $a(r)=0$. Applying a similar treatment to the scalar's field equation, it is then easy to see that the only possibility is $\left|\phi\right|=v$, proving that the Abrikosov-BTZ black hole will always be hairless.

As discussed in \cite{lee}, the reason why monopoles escape the no-hair theorems can be tracked down to the effective potential of the theory having a position-dependent absolute minimum. At large $r$, the minimum roughly corresponds to the symmetry-broken phase with a radial magnetic field, whereas at smaller $r$, this privilege shifts back to the symmetric phase. Thus, both the core and the outskirts of the 'tHP monopole are at the absolute minimum. In the presence of a horizon, the exterior matter fields will take the configuration that is energetically most favorable. So, for a small-enough horizon, they will reproduce the non-trivial configuration of the monopole tail, generating what we understand as a 'tHP monopole holding a BH within its core.

The situation is qualitatively different for Abrikosov vortices. From \eqref{eq:curvedaction2}, we can read off the effective potential
\begin{equation}
	U\left(a,\left|\phi\right|\right) = \frac{a^2}{r^2}\left|\phi\right|^2 + \frac{\lambda}{4}\left(\left|\phi\right|^2-v^2\right)^2\;.
	\label{eq:effective_potential}
\end{equation}
This potential has a ``saddle line'' of degenerate extrema at $\left|\phi\right|=0,\, \forall a$, but its absolute minimum is at the symmetry-broken phase, $\left|\phi\right|=v, \, a=0$, for any $r$. Thus, while the outskirts of the Abrikosov vortex are at the minimum of the potential, its core lies on a saddle. What holds the core there is solely the regularity condition \eqref{eq:flat_origin_condition}. But as soon as a horizon is introduced, independently of its size, this condition gets dropped and the fields can relax to the true minimum; their asymptotic values. The essential difference between \eqref{eq:effective_potential} and the effective potential for 'tHP monopoles is an additional term in the latter deriving from the intrinsically non-Abelian part of the Yang-Mills action of \eqref{eq:SU(2)_theory}. We can therefore blame the Abelian nature of Abrikosov vortices for the lack of BTZ vortex hair. It is amusing to speculate that non-Abelian vortices might provide hair for BTZ black holes.

In the above analysis, we were mostly concerned about the behavior outside the horizon. However, by the regularity condition at the horizon, the solution can be continued to the interior, where the matter fields remain constant. The full solution in the presence of a horizon is, therefore, the usual (hairless) non-rotating BTZ black hole pierced by a point solenoid of magnetic flux $2\pi N$ in the symmetry-broken phase,
\begin{align}
	&\quad\phi = v e^{iN\theta}\,, \qquad\qquad eA_\mu dx^\mu = -Nd\theta\,, \nonumber \\
	ds^2 &= \left(-M+\frac{r^2}{l^2}\right)dt^2 - \frac{1}{\left(-M+\frac{r^2}{l^2}\right)}dr^2 -r^2 \,d\theta^2 \;,
	\label{eq:solenoid_BTZ}
\end{align}
whose horizon is at $r_h=\sqrt{M}\ell$. Note that these matter fields are ill-defined at the origin, but this is not worrisome since this point is hidden behind the horizon. In a sense, this BH is the (2+1)-dimensional analogue of the magnetic RN black hole, understanding point solenoids as the (2+1)-dimensional analogues of Dirac monopoles.

During the completion of this work, \cite{ghosh} appeared on the arXiv, where they studied black holes in the same Abrikosov (a.k.a.\ Nielsen-Olesen) vortices. Their main result was that such BHs would have a different temperature from the BTZ black hole while sharing the same $r_h$, which seemed puzzling. However, their analysis focused mostly on the metric rather than the matter fields and their result was expressed in terms of generic $\rho(r)$ and $a(r)$. In our notation, their result reads
\begin{equation}
	T= \exp \left[\int_0^{r_h} dr\left(C_1 r\frac{\left(\rho'\right)^2}{\rho} + C_2 \frac{\left(a'\right)^2}{r}\right)\right]T_{\text{BTZ}}\;,
\end{equation}
where $C_1$, $C_2$ are constants that will be unimportant for our discussion. Note that the difference between temperatures depends exclusively on the derivatives of the matter fields inside the horizon. We have argued that Abrikosov vortices in the presence of a horizon do not escape the no-hair theorems and they always reduce to the hairless magnetic BTZ black hole \eqref{eq:solenoid_BTZ}. So this seems to resolve the temperature puzzle in \cite{ghosh}; the matter fields inside the horizon are constant and hence the temperatures of the usual BTZ and the Abrikosov-BTZ black holes remain the same.

\section{Vortices in AdS/CFT} \label{sec:ads/cft}
The AdS$_{d+1}$/CFT$_d$ correspondence establishes a duality between asymptotically AdS gravitational theories in $d+1$ dimensions and conformal field theories (CFT) at the $d$-dimensional boundary (see e.g.\ \cite{MAGOO} and references therein). Given that the vortex solutions described in the previous section live in an asymptotically AdS$_3$ space-time, they are naturally covered by this correspondence and must have a CFT$_2$ interpretation. Here, we discuss (after a brief review of AdS/CFT) some features of these bulk solutions in the context of this correspondence and provide some insight toward their CFT$_2$ interpretation.

The AdS/CFT correspondence is twofold. First, there is an equivalence between the Hilbert spaces of the bulk and boundary theories. That is, every state in the bulk theory corresponds to a state in the boundary theory that, by the state-operator map of CFTs, is associated to a local operator. Second, bulk solutions that do not decay fast enough near the boundary are interpreted as perturbations of the boundary theory. So, given a solution of the bulk field equations, the first question should be whether it is dual to a state or a perturbation. The distinction between these cases is closely related to the notion of normalizability \cite{balasubramanian} since a bulk solution must be normalizable --with respect to a given inner product-- in order for it to be a state of the Hilbert space.

The normalizability of a bulk solution is determined from its decay near the boundary \cite{balasubramanian}. To be definite, consider the asymptotic behavior of a scalar field of mass $m$ in AdS$_{d+1}$,
\begin{equation}
	\psi\to \frac{\alpha(x)}{r^{\Delta_-}} + \frac{\beta(x)}{r^{\Delta_+}}\;,
	\label{eq:alpha_beta_decays}
\end{equation}
where $x$ denotes the dependence on the coordinates of the $d$-dimensional boundary and
\begin{equation}
	\Delta_\pm = \frac{d}{2}\pm \sqrt{\frac{d^2}{4}+m^2\ell^2}\;.
	\label{eq:scalar_delta}
\end{equation}
In general, the $\beta$-term is normalizable while the $\alpha$-term is not, and $\psi$ is then said to be dual to a primary operator $\mathcal{O}$ of dimension $\Delta_+$. A solution with $\alpha\neq 0$ is non-normalizable and corresponds to a deformation of the CFT by a term
$$\int d^dx\, \alpha(x)\mathcal{O}\;,$$
and so the coefficient $\alpha$ is interpreted as a source. Then, the coefficient of the subleading term, $\beta$, corresponds to the vacuum expectation value of the dual operator in the presence of the source, $\beta=\left<\mathcal{O}\right>_\alpha$.

In contrast, a solution with $\alpha=0$ is normalizable and describes a state $\left|\Psi\right>$ of the Hilbert space. In this case, $\beta$ is interpreted as the expectation value of $\mathcal{O}$ on the corresponding CFT state \cite{kraus}, $\beta=\left<\Psi\right|\mathcal{O}\left|\Psi\right>$. In the range $-d^2/4<m^2\ell^2<1-d^2/4$ just above the BF bound, however, both terms in \eqref{eq:alpha_beta_decays} are normalizable and another quantization is possible, giving a dual operator of dimension $\Delta_-$. In this quantization, $\alpha$ and $\beta$ exchange their roles as source and expectation value. A similar story holds for bulk vector fields of mass $m$ \cite{MAGOO}, which may be dual to vector operators of dimensions
\begin{equation}
		\Delta_\pm = \frac{d}{2}\pm \sqrt{\frac{(d-2)^2}{4}+m^2\ell^2}\;.
		\label{eq:vector_delta}
\end{equation}

The bulk theory \eqref{eq:curvedaction} initially comprises a charged scalar $\phi$ and a gauge field $A_\mu$ that are dual, respectively, to a scalar operator and a conserved current in the CFT. But vortex solutions belong to the symmetry-broken phase, where the gauge field becomes massive via the Higgs mechanism. Thus, the matter content near the boundary consists in a real scalar field of mass $m_s^2=\lambda v^2$ and a massive vector field of mass $m_A^2=2e^2v^2$ that, by the AdS/CFT dictionary (\ref{eq:scalar_delta},\ref{eq:vector_delta}), are dual to a scalar operator $\mathcal{O}$ and a vector operator $\mathcal{W}_\mu$ of scaling dimensions
\begin{equation}
	\Delta_{\mathcal{O}}= 1 \pm \sqrt{1+\lambda v^2\ell^2}\,, \qquad \Delta_{\mathcal{W}}= 1 + \sqrt{2}ev\ell\;.
	\label{eq:dual_ops}
\end{equation}
The $\Delta_-$ quantization for $\mathcal{O}$ is only allowed when $-1<\lambda v^2 \ell^2<0$, whereas for $\mathcal{W}_\mu$ it is not allowed at all since it would break unitarity. In the CFT side, the Higgs mechanism is realized as multiplet recombination (see e.g.\ \cite{bianchi,tachikawa}), by which a conserved current can ``eat'' a marginal scalar operator and acquire an anomalous dimension that breaks its conservation. From \eqref{eq:scalar_delta}, a scalar operator with $\Delta=d$ is dual to a massless scalar field in the bulk, so this quite literally reproduces how the gauge field in the bulk ``eats'' a (massless) Goldstone boson to acquire a mass that breaks the symmetry.

Clearly, by the asymptotic behavior \eqref{eq:ads_decay}, Abrikosov vortices are normalizable solutions. Indeed, the gauge field in Cartesian coordinates is proportional to $\sim a(r)/r$, which decays as $\sim r^{-\Delta_{\mathcal{W}}}$, and the radial field goes as $\sim r^{-\Delta_{\mathcal{O}}}$. Therefore, a vortex solution forms a state of the Hilbert space that is dual to some state $\left|\psi\right>$ of the CFT\footnote{As discussed in \cite{bolognesi} for AdS$_4$ monopoles, in the case of negative $\lambda$ we should expect two different vortex solutions with the different decays of \eqref{eq:ads_decay}. Each of these would form a state in the corresponding quantization. But like in the sections above, we focus our discussion on the case $\lambda>0$.}. Like $\beta$ from \eqref{eq:alpha_beta_decays} in the case of $\alpha=0$, the coefficients of these decaying terms correspond to the expectation values of the dual operators on the state $\left|\psi\right>$. Schematically,
\begin{equation}
	\left|\phi\right| \to v + \frac{\left<\psi\right|\mathcal{O}\left|\psi\right>}{r^{\Delta_\mathcal{O}}}\,, \qquad eA_\mu \to -N\partial_\mu\theta + \frac{\left<\psi\right|\mathcal{W}_\mu\left|\psi\right>}{r^{\Delta_\mathcal{W}}}\;.
\end{equation}
Although we will not compute these coefficients, we note that they are non-zero for the non-singular solutions in Fig.~\ref{fig:asymads_matter} but they vanish when a horizon is turned on since the fields take a trivial configuration. The question is then, what kind of state are vortices dual to?

We know that vortex solutions are characterized by their vorticity $N\in \mathbb{Z}$, so there should exist some topological invariant that classifies their dual states. The topology of the CFT$_2$ is obtained from the limit to the boundary of global AdS$_3$ \eqref{eq:adsmetric} and it consists in the cylinder $\mathbb{R}^1_t\times S_\theta^1$. In this topology, states can separate in different so-called winding sectors. To see this, we can consider preparing states of the CFT as in \cite{DSD} by cutting open the path integral of a hypothetical Lagrangian description of the theory. With a compact fundamental field (e.g.\ $\varphi\sim \varphi + 2\pi$), states would fall in distinct sectors where we sum only over configurations in which $\varphi$ winds $N$ times around the $S_\theta^1$ (i.e.\ $\varphi(t,\theta=2\pi)=\varphi(t,0)+2\pi N$) and they would therefore be labelled by their winding number $N$. The topological nature of this parameter suggests that we identify it with the vorticity in the bulk, so that vortex solutions of vorticity $N$ are dual to CFT$_2$ states in the sector of winding $N$.

Furthermore, the simplicity of our solutions (minimal energy, static and radial symmetry) suggests that we identify them with a fundamental state $\left|N\right>$ of that sector. Fluctuations around the bulk vortex solutions would then correspond to other states in the same sector obtained by acting with local operators on top of $\left|N\right>$. By the state-operator map \cite{DSD}, $\left|N\right>=\sigma_N(0)\left|0\right>$, we would conclude that vortices are dual to a winding operator $\sigma_N$ that, similarly to the twist operator from \cite{ginsparg}, is attached to a topological line that makes $\varphi$ wind $N$ times upon crossing. These expectations are in line with \cite{bolognesi}, where it is stated that asymptotically AdS$_4$ 'tHP monopoles are dual to monopole operators in the CFT$_3$ \cite{borokhov}. These operators are topological disorder operators that generalize the 2-dimensional winding operator to $d=3$. Thus, this would materialize a CFT version of the statement that 'tHP monopoles are the 4-dimensional generalization of $D=2+1$ Abrikosov vortices, invoked in several occasions throughout this paper.

An important caveat to the precise identification of the CFT$_2$ dual is that the bulk theory \eqref{eq:curvedaction} has a trivially conserved current,
\begin{equation}
	\tilde{J}^\mu = -e\epsilon^{\mu\nu\rho}F_{\nu\rho}\;,
	\label{eq:hidden_current}
\end{equation}
which describes an additional $U(1)$ global symmetry. Its conservation is equivalent to the Bianchi identity and its conserved charge is nothing but the magnetic flux\footnote{In curved space, we define $\tilde{J}^\mu = -e\frac{\epsilon^{\mu\nu\rho}}{\sqrt{g}}F_{\nu\rho}$ and the conservation reads $\partial_\mu(\sqrt{g}\tilde{J}^\mu)=0$. But the conserved charge still corresponds to the magnetic flux,
\begin{equation*}
	\int_{\Sigma_t} d^2x\, \sqrt{g}\tilde{J}^0 = -e\int_{\Sigma_t} \mathbf{F}_{(2)} = \Phi\;.
\end{equation*}
The awkward minus sign comes from the flat-space definition $B=F_{21}$, which identifies $B$ with the $z$ component of a 3-dimensional magnetic field crossing the 2-dimensional spatial sheet, in mostly negative signature.},
\begin{equation}
	\int d^2x\, \tilde{J}^0 = e\int d^2x\, B = \Phi\;.
\end{equation}
It is widely believed that in any consistent theory of quantum gravity there should be no bulk global symmetries \cite{banks}, so it does not look like the Abrikosov vortex is fully consistent in AdS/CFT as is. Presumably, the bulk theory should be modified so that this symmetry is gauged.

A way to achieve this would be to add a Chern-Simons (CS) term,
\begin{equation}
	\frac{k}{4\pi}\int d^3x\, \epsilon^{\mu\nu\rho}A_\mu\partial_\nu A_\rho\;,
\end{equation}
to \eqref{eq:curvedaction}. This would help out in the identification of the CFT$_2$ dual because, in $d=2$, it is actually the CS term what connects a bulk gauge field to a (chiral) conserved current of the CFT \cite{kraus_per,jensen}. However, this would yield CS vortices rather than Abrikosov vortices in the bulk \cite{dunne}. As opposed to Abrikosov vortices, CS vortices carry electric charge in addition to the magnetic flux and they therefore have spin. When coupled to gravity, this induces a non-zero angular momentum for the metric and the analysis becomes more complicated. It would be interesting to extend our analysis to these vortices and find out whether they can provide hair for the BTZ black hole. Another option to gauge \eqref{eq:hidden_current} would be to couple it to a new gauge field $B_\mu$ by adding a term like
\begin{equation}
	\frac{k}{2\pi}\int d^3x\, \epsilon^{\mu\nu\rho}B_\mu\partial_\nu A_\rho\;,
\end{equation}
which arises in certain compactifications of String Theory \cite{jensen}. In this theory, Abrikosov vortices would remain spinless as they would only carry $B$-electric charge in addition to the $A$-magnetic flux. But further studies would be needed to determine if $B_\mu$ should also couple to the scalar field and, in particular, if it should get a mass via the Higgs mechanism.

\subsection{Comparison with the holographic superconductor}
We conclude with a few remarks on the comparison between the solutions discussed above and the story of the holographic superconductor (see e.g.\ \cite{horowitz} for an introduction to the subject). Consider a scalar field charged under a $U(1)$ gauge symmetry in asymptotically AdS space, i.e.\ \eqref{eq:curvedaction} but with an ordinary mass instead of the symmetry-breaking potential. In this theory, an electrically charged BH can develop nontrivial scalar hair outside the horizon when its radius $r_h$ is small enough \cite{gubser}. Qualitatively, the reason is that the electric potential $A_0$ gives an effective negative mass squared to the scalar field near the horizon that destabilizes it and makes it condense. From the CFT point of view, this corresponds to a state on which the operator $\mathcal{O}$, dual to $\phi$, has a non-zero expectation value related to its decay (recall \eqref{eq:alpha_beta_decays}). But when $r_h$ is increased beyond a critical value, the BH becomes hairless with trivial $\phi$ outside the horizon, leaving $\left<\mathcal{O}\right>=0$.

Black holes are thermal states with a temperature $T\sim r_h$ given by the Hawking temperature. Thus, the hair/no-hair transition in the bulk describes a second-order phase transition driven by temperature at the boundary. At large $T$, the system is in the symmetric phase $\left<\mathcal{O}\right>=0$, but below a critical temperature $T_c$, the scalar operator acquires an expectation value $\left<\mathcal{O}\right>\neq 0$ that spontaneously breaks the $U(1)$ symmetry. This is precisely the sort of phase transitions observed in superconductors, so this setup provides the means for studying a superconductor from a higher-dimensional (gravitational) system, hence the name {\it holographic superconductor}\footnote{In general, a gauge symmetry in the bulk corresponds to a global symmetry at the boundary, so one should actually talk about a superfluid rather than a superconductor \cite{horowitz}. However, in \cite{domenech} it was shown (for $D=3+1$) that one can get a proper superconductor by imposing Neumann boundary conditions on the bulk gauge field since that yields a $U(1)$ {\it gauge} symmetry at the boundary. This allows one to construct Abrikosov vortices at the boundary (as in \cite{zeng}). But as we are imposing Dirichlet boundary conditions (recall \eqref{eq:ads_bdy_cond}) and working in $D=2+1$ to get an Abrikosov vortex in the bulk, this does not apply to our configuration and we indeed ought to talk about a superfluid rather than a superconductor.} (HS).

The bulk side of this story is somewhat similar to the phenomenology of the solutions from Section \ref{sec:asym_ads}, so we expect the CFT dual of our solutions to have an interpretation on the lines of the HS. Yet, there are some differences between these systems that we now outline. First, the bulk fields in our case only have a non-trivial decay toward the boundary in non-singular solutions (recall Fig.~\ref{fig:asymads_matter}). As soon as a horizon is turned on, the matter fields jump to their asymptotic values and make the BH hairless. In a sense, we can say that the black hole acquires hair only when $r_h=0$. Thus, on the CFT side, this seems to translate into a phase transition with critical temperature $T_c=0$, which might not be too surprising given the low dimensionality of the boundary theory (i.e.\ $d=2$).

Second, what induces the ``hair'' in our solutions is not an electric charge but a magnetic flux piercing the black hole. In the story of the HS, this would imply that the superconducting phase is conditioned by a background magnetic field rather than a chemical potential \cite{horowitz}. And third, vortex solutions are labelled by their vorticity, so we expect the dual system to involve some topological invariant. This hints at a topological phase transition in the boundary theory when the vorticity of the bulk changes. A well-known system showing these last two features is, in fact, still a superconductor but in the shape of a hollow cylinder with a coaxial magnetic field \cite{arutunian}. In such a system, the supercurrent winds around the cylinder due to the magnetic field and it can only take quantized values corresponding to the winding of the order parameter. As the magnetic field is continuously changed, the superconductor changes between states of different winding in sudden jumps, as was observed in the '60s by Little and Parks \cite{LittleParks}.

We find this system very suggestive and we believe that it might correspond to the CFT dual of our solutions. That is, a (1+1)-dimensional cylindrical superconductor with a coaxial magnetic flux where the quantized values of the supercurrent are identified with the vorticity in the bulk. Thus, magnetic vortices in the bulk seem to expand our understanding of the HS by allowing us to probe holographically how cylindrical superconductors react to external magnetic fields. This was studied for a (2+1)-dimensional cylindrical HS in \cite{montull,pujolas} by adding a vortex line to an electrically charged string in the $D=3+1$ bulk, in harmony with our argument. However, in those works they neglected the bulk scalar's potential, which allowed them to take the boundary condition $|\phi|\to 0$, whereas in our case the scalar's potential plays an essential role in giving the vortex topological stability via the boundary condition $|\phi|\to v$.

We should not end without pointing out that this sets an important drawback to our interpretation. In contrast with the usual HS, our bulk solutions are asymptotically in the symmetry-broken phase and, as discussed around $\eqref{eq:dual_ops}$, in the CFT side this translates into a multiplet recombination that breaks the $U(1)$ symmetry. Thus, in our case, a non-vanishing $\left<\mathcal{O}\right>$ does not break the symmetry spontaneously since it is already broken and the hair/no-hair transition no longer seems to translate directly into a superconducting phase transition. Although a more detailed study addressing this and other issues is undoubtedly needed, we believe that our interpretation opens up interesting lines of investigation for the HS. For example, studying whether purely magnetic vortex lines can hold a black string at their core might shed light on new properties of (2+1)-dimensional HSs.

\paragraph{Note added.} After submitting this work to the arXiv, we were made aware of \cite{edery}, where they had studied the same problem as us albeit with the focus on a different aspect of it. In particular, they centered their analysis on non-singular vortex solutions and they presented different ways of computing their mass. Our discussion for non-singular solutions is in agreement with their results and our work extends their analysis to singular solutions and some aspects of AdS/CFT.

\acknowledgments
I would like to thank Jennifer Cano, Gabriel Cuomo, Roberto Emparan, Leonardo Rastelli and Luigi Tizzano for very useful discussions and Martin Ro\v{c}ek for his helpful comments on the manuscript. I would also especially like to thank Zohar Komargodski for his help and guidance throughout this project as well as for his suggestions for the manuscript. This work was supported in part by
NSF grant \# PHY-1915093.

\bibliographystyle{ytphys}
\bibliography{references_abrik_paper}

\end{document}